\documentclass[fleqn,usenatbib]{mnras}

\usepackage{newtxtext,newtxmath}

\usepackage[T1]{fontenc}

\DeclareRobustCommand{\VAN}[3]{#2}
\let\VANthebibliography\thebibliography
\def\thebibliography{\DeclareRobustCommand{\VAN}[3]{##3}\VANthebibliography}

\usepackage{hyperref,enumerate}
\usepackage{graphicx}	
\usepackage{amsmath}	
\usepackage{multirow}
\usepackage{cuted}
\usepackage{longtable}
\usepackage{soul}
\usepackage[T1]{fontenc}
\usepackage{gensymb}
\usepackage{ulem}
\usepackage{verbatim}

\title[A multiplicative modulator for the bispectrum in BHXRBs]{A possible connection between the ridge and hypotenuse in the bispectrum of black hole X-ray binaries through quasi-periodic modulation}

\author[P.\ Jin et al.]{
Pei~Jin,$^{1}$
Mariano~M\'{e}ndez,$^{1}$
Federico~Garc\'{\i}a,$^{2}$
Diego~Altamirano,$^{3}$
Haifan~Zhu,$^{1}$
Liang~Zhang,$^{4,5}$
\newauthor{and Guobao Zhang$^{5,6}$}
\\
$^{1}$Kapteyn Astronomical Institute, University of Groningen, P.O.\ BOX 800, 9700 AV Groningen, The Netherlands\\
$^{2}$Instituto Argentino de Radioastronom\'{\i}a (CCT La Plata, CONICET; CICPBA; UNLP), C.C.5, (1894) Villa Elisa, Buenos Aires, Argentina\\
$^{3}$ School of Physics and Astronomy, University of Southampton, Southampton, Hampshire SO17 1BJ, UK\\
$^{4}$ Key Laboratory of Particle Astrophysics, Institute of High Energy Physics, Chinese Academy of Sciences, Beijing 100049, People’s
Republic of China\\
$^{5}$University of Chinese Academy of Sciences, Beijing 100049, People's Republic of China\\
$^{6}$Yunnan Observatories, Chinese Academy of Sciences, Kunming 650216, People's Republic of China
}

\date{}

\pubyear{2026}

\begin{document}
\label{firstpage}
\pagerange{\pageref{firstpage}--\pageref{lastpage}}
\maketitle

\begin{abstract}

Quasi-periodic oscillations (QPOs) in black-hole X-ray binaries show nonlinear coupling with broadband variability, producing characteristic ridge, hypotenuse, and web-like structures in the bicoherence. However, the origin of these structures remains unclear.
We develop a mathematical model in which a modulator phase-locked to the QPO acts multiplicatively on the underlying variability.
The resulting frequency mixing naturally produces both the ridge and hypotenuse associated with the fundamental QPO, as two parts of the same modulation-induced bispectral structure, and the same mechanism extends to the second harmonic. 
The model further predicts opposite-sign biphases between each ridge and its corresponding hypotenuse. 
We test these predictions using NICER observations of Swift~J1727.8$-$1613 and MAXI~J1820$+$070. 
The bispectral maps show QPO-related ridge--hypotenuse structures broadly consistent with the model, with qualitatively opposite-sign biphases between each ridge and its corresponding hypotenuse. In particular, the biphase map provides a complementary diagnostic that reveals weak coupling structures that are difficult to identify from the bicoherence alone. 
We also compare the measured bispectral modulus with the predictions for isolated ridges and hypotenuses. These predictions broadly reproduce the observed
frequency dependence, while localized deviations remain near the fundamental-QPO and second-harmonic frequencies, where different bispectral structures overlap. 
Our results provide a simple mathematical framework in which the web-like pattern in the bispectrum could arise naturally from multiplicative coupling between a QPO-phase-locked modulator and the underlying variability.

\end{abstract}




\section{INTRODUCTION}
\label{sec:introduction}

Black-hole X-ray binaries (BHXBs) exhibit significant short-timescale variability ranging from milliseconds to hundreds of seconds~\citep[see reviews by ][and references therein]{2011BASI...39..409B, 2019NewAR..8501524I, 2025Galax..13..111Z}.
As an outburst of one of these sources evolves, the properties of the short-timescale variability also change, including their characteristic frequencies~\citep[e.g.,][]{2005A&A...440..207B, 2011MNRAS.418.2292M, 2020MNRAS.494.1375Z, 2025MNRAS.538.1143L, 2025A&A...699A...9J} and strength~\citep[e.g.,][]{1997ApJ...479..926M, 2011MNRAS.410..679M}. These changes provide a powerful way to trace the evolution of the physical and geometrical properties of the accretion flow throughout an outburst.

Over the past two decades, the power density spectrum (PDS) has been widely used to study the short time-scale variability of BHXBs. One of the most prominent features in the PDS are the quasi-periodic oscillations (QPO), which appear as a relatively narrow peak above the broadband variability~\citep{1999ApJ...526L..33W, 2002ApJ...564..962R, 2005ApJ...629..403C}. The PDS is commonly modelled as a linear combination of Lorentzian components, with different components used to describe the QPOs and broadband variability~\citep[e.g.,][]{2000MNRAS.318..361N, 2002ApJ...572..392B}. In such a decomposition, the different variability components are usually treated as independent components, without considering possible coupling between them.

The cross spectrum (CS) characterizes the relation between concurrent light curves in two energy bands at the same Fourier frequency~\citep{1996MNRAS.280..227N, Bendat-2010}. Compared with the PDS, which contains only information about the variability amplitude, the CS introduces an additional dimension, namely the Fourier phase lag between the two light curves at each Fourier frequency. This additional information can reveal variability components that are blended and therefore difficult to distinguish in the PDS alone~\citep{2024MNRAS.527.9405M, 2024A&A...687A.284K, 2025A&A...696A.128B, 2025A&A...696A.237F, 2025arXiv250507938B, 2025ApJ...990...43R, 2026A&A...706A.208J}.

Both the PDS and CS describe variability at individual Fourier frequencies and do not directly measure the coupling between different Fourier frequencies. However, there is evidence that variability at different Fourier frequencies is not independent. First, the type-C QPO and strong broadband variability evolve closely together and can disappear simultaneously during the transition from the hard to the soft state~\citep[e.g.,][]{2020ApJ...891L..29H, 2026A&A...706A.208J}, suggesting a close connection between these variability components. 
More direct evidence for coupling between the QPO and broadband variability comes from bicoherence measurements~\citep[e.g.,][]{2011MNRAS.413.1819M, 2020MNRAS.491..313A, 2022MNRAS.514.1720A, 2024MNRAS.529.4602Z, 2024ApJ...974..303Z}.
The bicoherence measures phase coupling among three Fourier frequencies, providing information about the nonlinear coupling between different variability components~\citep{4317207}.
These measurements reveal three characteristic bicoherence patterns, commonly referred to as the ``web'', ``hypotenuse'', and ``cross'' (characterized by vertical and horizontal ridges)  patterns~\citep[e.g.,][]{2011MNRAS.413.1819M, 2020MNRAS.491..313A}.

Two physical scenarios have been proposed to qualitatively explain the phase coupling between the QPO and broadband variability. In the propagating-fluctuation model~\citep{1997MNRAS.292..679L, 2006MNRAS.367..801A, 2013MNRAS.434.1476I}, slow mass-accretion-rate fluctuations generated at large radii propagate inwards and modulate variability produced at smaller radii.
Alternatively, the physical or geometrical properties of the region that produces the QPO may vary periodically with the QPO. If these variations change the response of this region to the underlying broadband variability, the QPO can modulate the broadband variability, naturally producing phase coupling between them~\citep{2020MNRAS.491..313A}.

In this work, we develop a mathematical model to describe the bispectral coupling between the QPO and the underlying variability.
We show that a modulator phase-locked to the QPO, acting multiplicatively on the variability, naturally produces both the ridge and hypotenuse structures in the bispectrum. In Section~\ref{sec:MATHEMATICAL FORMALISM} we introduce the mathematical model and derive its main predictions. 
In Section~\ref{sec:DATA ANALYSIS AND RESULTS} we present the bispectral analysis and compare the observations with the model predictions. 
In Section~\ref{sec:DISCUSSION} we discuss the implications of our results, and in Section~\ref{sec:CONCLUSIONS} we summarize our main conclusions.

\section{MATHEMATICAL FORMALISM}
\label{sec:MATHEMATICAL FORMALISM}

\subsection{Definitions}
\label{subsec:definitions}

For a light curve \(x(t)\), we denote its Fourier transform by \(X(\nu)\). The bispectrum at Fourier frequencies $\nu_1$ and $\nu_2$ is defined as
\begin{equation}
B(\nu_1,\nu_2)
=
\left\langle
X(\nu_1)X(\nu_2)X^*(\nu_1+\nu_2)
\right\rangle ,
\label{eq:bispectrum_def}
\end{equation}
where the asterisk means complex conjugate, and the average is taken over data segments~\citep{2002MNRAS.336..817M}. 
The modulus of the bispectrum is denoted by $|B|$, while the biphase is the phase of the bispectrum,
\begin{equation}
\phi_{\rm bi}(\nu_1,\nu_2)
=
\arg B(\nu_1,\nu_2),
\label{eq:biphase_def}
\end{equation}
The squared bicoherence~\citep{4317207} is defined as
\begin{equation}
b^2(\nu_1,\nu_2)
=
\frac{
\left|
\left\langle
X(\nu_1)X(\nu_2)X^*(\nu_1+\nu_2)
\right\rangle
\right|^2
}{
\left\langle |X(\nu_1)X(\nu_2)|^2 \right\rangle
\left\langle |X(\nu_1+\nu_2)|^2 \right\rangle
}.
\label{eq:bicoherence_def}
\end{equation}

\subsection{A multiplicative modulator}
\label{subsec:model}

Assuming that a quasi-periodic modulator $m(t)$ multiplicatively modulates all
underlying timing variability components, we write
\begin{equation}
\left\{
\begin{aligned}
x(t)
&=
\left[1+m(t)\right]s(t), \\
s(t)
&= q(t)+\sum_j l_j(t).
\end{aligned}
\right.
\label{eq:model_basic}
\end{equation}
Here, $q(t)$ represents the fundamental QPO and its second harmonic, and each $l_j(t)$ represents an additional intrinsic variability component.
Following~\citet{2024MNRAS.527.9405M}, we assume that these variability processes are Lorentzian functions that are mutually incoherent.
We further assume that the variability components represented by the different Lorentzians are independent Gaussian stochastic processes. 
Under these assumptions, the intrinsic bispectrum of each of these components, as well as mixed third-order terms involving different components, vanish under ensemble averaging.

We allow one exception: the fundamental QPO and its harmonic could be phase-coupled. 
Because they are well separated in frequency, such coupling need not be apparent in second-order statistics, including the power spectrum and cross spectrum, but it can contribute directly to the bispectrum, as discussed in Section~\ref{subsec:intrinsic_qpo_harmonic}.

We assume \(m(t)\) is a quasi-periodic modulator at the QPO frequency and approximate it by its fundamental and second harmonic,
\begin{equation}
m(t)
=
u_1 \cos(2\pi \nu_0 t+\psi_{m,1})
+
u_2 \cos(4\pi \nu_0 t+\psi_{m,2}),
\label{eq:m_time}
\end{equation}
where \(\nu_0\) is the QPO fundamental frequency, and \(\psi_{m,1}\) and \(\psi_{m,2}\) are the phases of the fundamental and second harmonic components of the modulator, respectively.
Similarly, we write the QPO component as
\begin{equation}
q(t)
=
a_1 \cos(2\pi \nu_0 t+\psi_1)
+
a_2 \cos(4\pi \nu_0 t+\psi_2),
\label{eq:q_time}
\end{equation}
where \(\nu_0\) is the QPO fundamental frequency, and \(\psi_{1}\) and \(\psi_{2}\) are the phases of the fundamental and second harmonic components of the QPO, respectively.
While this simple approximation neglects the fact that QPOs have a finite width, it is sufficient to capture the basic frequency-mixing behaviour.
We assume that the fundamental and second harmonic of the modulator are phase-locked to the corresponding QPO components, and define the phase offsets as
\begin{equation}
\psi_{m,1}=\psi_1+\Delta_1,
\qquad
\psi_{m,2}=\psi_2+\Delta_2.
\label{eq:delta_def}
\end{equation}

The Fourier transform of $x(t)$ is then
\begin{align}
X(\nu)
={}& S(\nu)
+\frac{u_1}{2}e^{i\psi_{m,1}}S(\nu-\nu_0)
+\frac{u_1}{2}e^{-i\psi_{m,1}}S(\nu+\nu_0)
\nonumber\\
&+\frac{u_2}{2}e^{i\psi_{m,2}}S(\nu-2\nu_0)
+\frac{u_2}{2}e^{-i\psi_{m,2}}S(\nu+2\nu_0),
\label{eq:x_fourier_shifted}
\end{align}
where $S(\nu)$ is the Fourier transform of $s(t)$. Thus, the modulation
copies Fourier components to sidebands separated by $\pm\nu_0$ and
$\pm2\nu_0$.

\subsection{Fundamental QPO ridge at \(\nu_1=\nu_0\)}
\label{subsec:qpo_ridge}

We first consider the ridge produced by the modulation at $\nu_0$,
\begin{equation}
B_{\rm ridge}(\nu_0,\nu)
=
\left\langle
X(\nu_0)X(\nu)X^*(\nu_0+\nu)
\right\rangle .
\label{eq:qpo_ridge_bispectrum}
\end{equation}
The key effect of the multiplicative modulation is that the intrinsic variability
\(s(t)\) is copied to sidebands separated by the modulation frequency \(\nu_0\).
This creates a phase-coupled contribution to the bispectrum,
\begin{equation}
B_{\rm ridge}^{m}(\nu_0,\nu)
\simeq
\frac{u_1a_1}{4}
e^{-i\Delta_1}
[\left\langle |S(\nu)|^2 \right\rangle + \left\langle |S(\nu + \nu_0)|^2 \right\rangle].
\label{eq:qpo_ridge_sum}
\end{equation}
For weak modulation, \(u_1 \ll 1\), we can approximate the intrinsic power spectrum of \(s(t)\) by the observed power spectrum \(P(\nu)\) of \(x(t)\), neglecting the effect of modulation on the power spectrum. The bispectral modulus then becomes \footnote{In this work we neglect second- and higher-order contributions arising from the modulation. We note that in this work the power spectrum at the real Fourier frequency is not multiplied by a factor of two.}
\begin{equation}
|B_{\rm ridge}(\nu_0,\nu)|
\simeq
\frac{u_1a_1}{4}
[P(\nu) + P(\nu + \nu_0)].
\label{eq:qpo_ridge_mag}
\end{equation}
We additionally introduce a phenomenological one-sided approximation in which the shift from higher to lower frequencies is suppressed. The bispectral modulus then becomes
\begin{equation}
|B_{\rm ridge}(\nu_0,\nu)|
\simeq
\frac{u_1a_1}{4}P(\nu).
\label{eq:qpo_ridge_mag_ltoh}
\end{equation}
The corresponding biphase is set by the phase offset between the fundamental QPO and the modulator fundamental,
\begin{equation}
\phi_{\rm bi}^{\rm ridge}(\nu_0,\nu)
\simeq
-\Delta_1 .
\label{eq:qpo_ridge_phase}
\end{equation}

\subsection{The hypotenuse coupling at \texorpdfstring{\(\nu_1+\nu_2=\nu_0\)}{}}
\label{subsec:qpo_hypotenuse}

In the previous section, we discarded the negative-frequency part of the Fourier transform of real-valued signals. This is because, when focusing on second-order statistics such as the power spectrum and the cross spectrum, the negative-frequency components do not contain independent information. However, these components enter the bispectrum through the frequency-sum constraint (\(\nu_1+\nu_2=\nu_0\)) and therefore contribute to the measured third-order phase coupling.

Therefore, in our model, the same modulation that produces the fundamental QPO ridge also predicts a bispectral feature satisfying\begin{equation}
\nu_1+\nu_2=\nu_0 .
\end{equation}
Along this line, usually called hypotenuse~\citep{2022MNRAS.514.1720A}, the bispectrum is defined as
\begin{equation}
B_{\rm hypot}(\nu_1,\nu_2)
=
\left\langle
X(\nu_1)X(\nu_2)X^*(\nu_0)
\right\rangle .
\label{eq:diagonal_bispectrum}
\end{equation}
The key point is that the modulation-shifted term in \(X(\nu_1)\) contains
\begin{equation}
X(\nu_1)
\supset
\frac{u_1}{2}e^{i\psi_{m,1}}S(\nu_1-\nu_0).
\end{equation}
Since along the hypotenuse \(\nu_1-\nu_0=-\nu_2\), and since the negative-frequency Fourier component satisfies \(S(-\nu_2)=S^*(\nu_2)\),
we expect a contribution to the bispectrum of the form
\begin{equation}
\frac{u_1 a_1}{4}e^{i\Delta_1}
\left\langle
S^*(\nu_2)
S(\nu_2)
\right\rangle .
\end{equation}
There is an analogous contribution in which the shifted term comes from \(X(\nu_2)\),
\begin{equation}
\frac{u_1 a_1}{4}e^{i\Delta_1}
\left\langle
S(\nu_1)
S^*(\nu_1)
\right\rangle .
\end{equation}
Hence, along the hypotenuse line, the bispectrum is
\begin{equation}
\begin{aligned}
B_{\rm hypot}(\nu_1,\nu_2)
&\simeq
\frac{u_1a_1}{4}
e^{i\Delta_1}
[\left\langle |S(\nu_1)|^2 \right\rangle + \left\langle |S(\nu_2)|^2 \right\rangle].
\end{aligned}
\label{eq:diagonal_result}
\end{equation}
For weak modulation, \(u_1 \ll 1\),
we expect that the modulus of the bispectrum along the hypotenuse line is
\begin{equation}
\begin{aligned}
|B_{\rm hypot}(\nu_1,\nu_2)|
&\simeq
\frac{u_1a_1}{4}
\left[P(\nu_1)+P(\nu_2)\right].
\end{aligned}
\label{eq:diagonal_mag}
\end{equation}
And the biphase is
\begin{equation}
\phi_{\rm bi}^{\rm hypot}(\nu_1,\nu_2)
\simeq
\Delta_1 .
\label{eq:qpo_hypotenuse_phase}
\end{equation}
We note that in this model the ridge and hypotenuse arise from the same modulation at \(\nu_0\) and represent two parts of the same two-dimensional bispectral structure.
Combining Eqs.~\ref{eq:qpo_ridge_phase} and \ref{eq:qpo_hypotenuse_phase}, the model gives the opposite-sign relation, 
\begin{equation}
\phi_{\rm bi}^{\rm ridge}(\nu_0,\nu)
\simeq
-
\phi_{\rm bi}^{\rm hypot}(\nu_1,\nu_2).
\label{eq:ridge_diagonal_phase_relation}
\end{equation}

\subsection{Extension to the second harmonic}
\label{subsec:second_harmonic}

The same arguments discussed before apply to the modulation at $2\nu_0$, which produces both a second-harmonic ridge and its corresponding hypotenuse.
Repeating the previous calculation gives the second-harmonic ridge prediction,
\begin{equation}
B_{\rm ridge}(2\nu_0,\nu)
\simeq
\frac{u_2a_2}{4}
e^{-i\Delta_2}
\left[
\left\langle |S(\nu)|^2 \right\rangle
+
\left\langle |S(\nu+2\nu_0)|^2 \right\rangle
\right].
\label{eq:harmonic_ridge_result}
\end{equation}
The second-harmonic ridge therefore has biphase,
\begin{equation}
\phi_{\rm bi}^{\rm ridge}(2\nu_0,\nu)
\simeq
-\Delta_2 .
\label{eq:harmonic_ridge_phase}
\end{equation}
Under the assumption \(u_2 \ll 1\), the modulus is expected to follow
\begin{equation}
|B_{\rm ridge}(2\nu_0,\nu)|
\simeq
\frac{u_2a_2}{4}
\left[
P(\nu)+P(\nu+2\nu_0)
\right].
\label{eq:harmonic_ridge_mag}
\end{equation}
Under the same one-sided approximation, retaining only the shift from lower to higher frequencies, the bispectral modulus becomes
\begin{equation}
|B_{\rm ridge}(2\nu_0,\nu)|
\simeq
\frac{u_2a_2}{4}
P(\nu).
\label{eq:harmonic_ridge_mag_ltoh}
\end{equation}

The corresponding hypotenuse feature satisfies
\begin{equation}
\nu_1+\nu_2=2\nu_0 .
\end{equation}
Using the corresponding negative-frequency component gives
\begin{equation}
B_{\rm hypot}(\nu_1,\nu_2)
\simeq
\frac{u_2a_2}{4}
e^{i\Delta_2}
\left[
\left\langle |S(\nu_1)|^2 \right\rangle
+
\left\langle |S(\nu_2)|^2 \right\rangle
\right],
\label{eq:harmonic_hypotenuse_result}
\end{equation}
with the biphase
\begin{equation}
\phi_{\rm bi}^{\rm hypot}(\nu_1,\nu_2)
\simeq
\Delta_2.
\label{eq:harmonic_hypotenuse_phase}
\end{equation}
The ridge and hypotenuse biphases therefore obey the same opposite-sign relation as for the fundamental,
\begin{equation}
\phi_{\rm bi}^{\rm ridge}(2\nu_0,\nu)
\simeq
-
\phi_{\rm bi}^{\rm hypot}(\nu_1,\nu_2).
\label{eq:harmonic_phase_relation}
\end{equation}
Under the assumption \(u_2 \ll 1\), the modulus is expected to follow
\begin{equation}
|B_{\rm hypot}(\nu_1,\nu_2)|
\simeq
\frac{u_2a_2}{4}
\left[
P(\nu_1)+P(\nu_2)
\right].
\label{eq:harmonic_hypo_mag}
\end{equation}

\subsection{Intersections of bispectral structures}
\label{subsec:intrinsic_qpo_harmonic}

The ridge and hypotenuse structures derived above intersect at several frequency pairs in the two-dimensional bispectrum. 
At such intersections, the corresponding contributions must be added as complex quantities before taking the bispectral modulus. 

In particular, $(\nu_0,\nu_0)$ lies at the intersection of the fundamental-QPO ridge and the second-harmonic hypotenuse. 
The contributions at this and other intersections, including $(\nu_0,2\nu_0)$ and $(2\nu_0,2\nu_0)$, are derived explicitly in Appendix~\ref{app:intersections}.
In addition to these modulation-induced contributions, the QPO fundamental and its second harmonic could be intrinsically phase-coupled. 
We parameterize this possible coupling by
\begin{equation}
\psi_2 = 2\psi_1 + \Delta_{12},
\end{equation}
where $\Delta_{12}$ is the intrinsic phase offset between the harmonic and twice the fundamental phase. If present, this coupling provides an additional contribution at $(\nu_0,\nu_0)$,
\begin{equation}
B_{\rm int}(\nu_0,\nu_0)
\simeq
\frac{a_1^2a_2}{8}e^{-i\Delta_{12}}.
\end{equation}

\section{DATA ANALYSIS AND RESULTS}
\label{sec:DATA ANALYSIS AND RESULTS}

In this section, we analyze phase coupling in three observations of Swift~J1727.8$-$1613 and one observation of MAXI~J1820$+$070 using bispectral techniques. We compare the measured bispectral properties with the predictions of the multiplicative-modulation model described above. The data were obtained with the Neutron star Interior Composition Explorer (NICER; \citealt{2016SPIE.9905E..1HG}) and processed using HEASOFT~v6.35 and CALDB~v20220331. We reprocessed the data with {\tt nicerl2} to produce cleaned event files. For Swift~J1727.8$-$1613, we use photons in the 0.5--12.0~keV energy range. For MAXI~J1820$+$070, we use the 1.5--12.0~keV range, excluding softer photons to improve the signal-to-noise ratio of the measured bispectrum.

For each observation, we divide the light curve into segments and calculate the Fourier transform of each 24~s segment. We then calculate the un-normalized bispectrum using Eq.~\ref{eq:bispectrum_def} and derive the bicoherence and biphase using Eq.s~\ref{eq:bicoherence_def} and \ref{eq:biphase_def}, respectively. We subtract the bicoherence bias following the procedure of~\citet{2002MNRAS.336..817M, 2022MNRAS.514.1720A}.
In Sections~\ref{subsec:bispectral-maps}, we plot the squared bicoherence, the bispectral modulus and the biphase maps without applying any additional rebinning.
In Sections~\ref{subsec:fundamental-ridge} and \ref{subsec:Second-harmonic ridge}, we calculate the bispectral modulus along the fundamental-QPO and
second-harmonic ridges by averaging the modulus across the ridges over the frequency intervals 
$[\nu_0-W_1\times\nu_0/2,\nu_0+W_1\times\nu_0/2]$ and
$[2\nu_0-W_2\times2\nu_0/2,2\nu_0+W_2\times2\nu_0/2]$, respectively,
where $W_1\times\nu_0$ and $W_2\times2\nu_0$ are the corresponding averaging widths.
We note that averaging the bispectral modulus, $\langle|B|\rangle$, is not equivalent to taking the modulus of the averaged complex bispectrum, $|\langle B\rangle|$. However, as shown in Appendix~\ref{lab:Averaging}, the two averaging procedures give very similar results over most of the frequency range considered here.
We adopt \(W_1=W_2=0.2\) throughout this analysis. 
Small changes in the averaging widths set by different choices of $W_1$ and $W_2$ do not affect the measured trends or our conclusions, although they change the associated uncertainties (see Appedix~\ref{lab:bispectral modulus for different Q}).
Similarly, in Sections~\ref{subsec:fundamental-hypotenuse} and \ref{subsec:harmonic-hypotenuse}, we calculate the hypotenuse bispectral modulus by averaging locally in the direction perpendicular to the hypotenuse line. The averaging width perpendicular to the hypotenuse line is set by the width of the fundamental QPO or the second harmonic, respectively.
Finally, we calculate un-normalized power spectra (without applying an additional factor of two, corresponding to the two-sided normalization; see~\citet{Bendat-2010}) to compare the bispectral modulus directly with the power spectrum.

In Sections~\ref{subsec:fundamental-ridge}, \ref{subsec:Second-harmonic ridge}, \ref{subsec:fundamental-hypotenuse}, and \ref{subsec:harmonic-hypotenuse}, we calculate three quantities for comparison with the bispectral modulus: (i) the Fourier-power triplet, given by the square root of the bicoherence denominator (Eq.~\ref{eq:bicoherence_def}); (ii) the prediction of the modulation model (Eqs.~\ref{eq:qpo_ridge_mag},~\ref{eq:diagonal_mag},~\ref{eq:harmonic_ridge_mag} and~\ref{eq:harmonic_hypo_mag}); and (iii) the power spectrum, corresponding to the phenomenological one-sided approximation (Eqs.~\ref{eq:qpo_ridge_mag_ltoh} and~\ref{eq:harmonic_ridge_mag_ltoh}) in which variability is shifted from lower to higher frequencies, while the reverse shift is suppressed.
The explicit expressions for these quantities are given in the corresponding subsections.
Since these quantities have different normalizations from the measured bispectral modulus, we introduce an independent scale factor for each comparison. The scale factor, A, is obtained by minimizing
\begin{equation}
\chi^2(A)=\sum_i\frac{\left(|B_i|-AF_i\right)^2}{\sigma_{|B|,i}^2+A^2\sigma_{F,i}^2},
\label{eq:scale_factor}
\end{equation}
where \(F_i\) is one of the quantities being compared with the measured bispectral modulus \(|B_i|\), and \(\sigma_{F,i}\) and \(\sigma_{|B|,i}\) are their corresponding uncertainties. The scale factor is fitted independently for each of the three quantities.
These comparisons are intended to test the frequency dependence of the bispectral modulus, rather than its absolute normalisation. The fitted scale factors are therefore treated as empirical normalisation constants and are not interpreted as measurements of the modulation amplitudes.
After obtaining the best-fitting scale factor \(A\), we plot the ratio \(|B_i|/(A F_i)\) to compare the frequency dependence of the measured bispectral modulus with that of each quantity.
Because these comparisons consider each ridge or hypotenuse contribution in isolation, residuals are expected at intersections where multiple bispectral structures contribute simultaneously. At $(\nu_0,\nu_0)$, $(\nu_0,2\nu_0)$, and $(2\nu_0,2\nu_0)$, the measured bispectrum is the complex sum of the relevant contributions, with the resulting modulus depending on both their amplitudes and relative phases, as discussed in Section~\ref{subsec:intrinsic_qpo_harmonic} and Appendix~\ref{app:intersections}.

\subsection{Bicoherence, bispectral modulus and biphase maps}
\label{subsec:bispectral-maps}

\subsubsection{Swift~J1727.8$-$1613}
\label{subsubsec:Swift~J1727.8$-$1613}

\begin{figure*}
    \centering
    \includegraphics[width=0.33\textwidth]{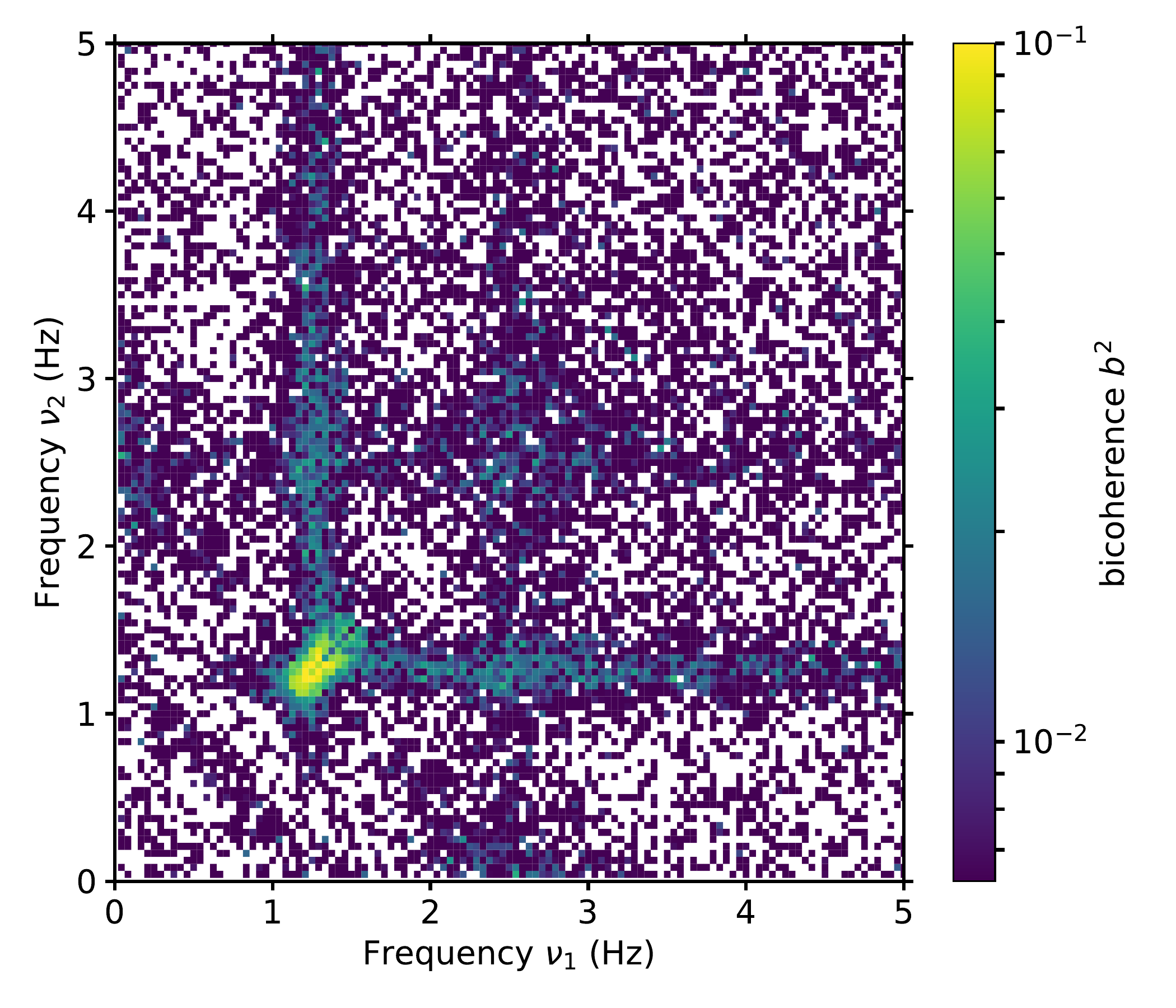}\hspace{-1mm}%
    \includegraphics[width=0.33\textwidth]{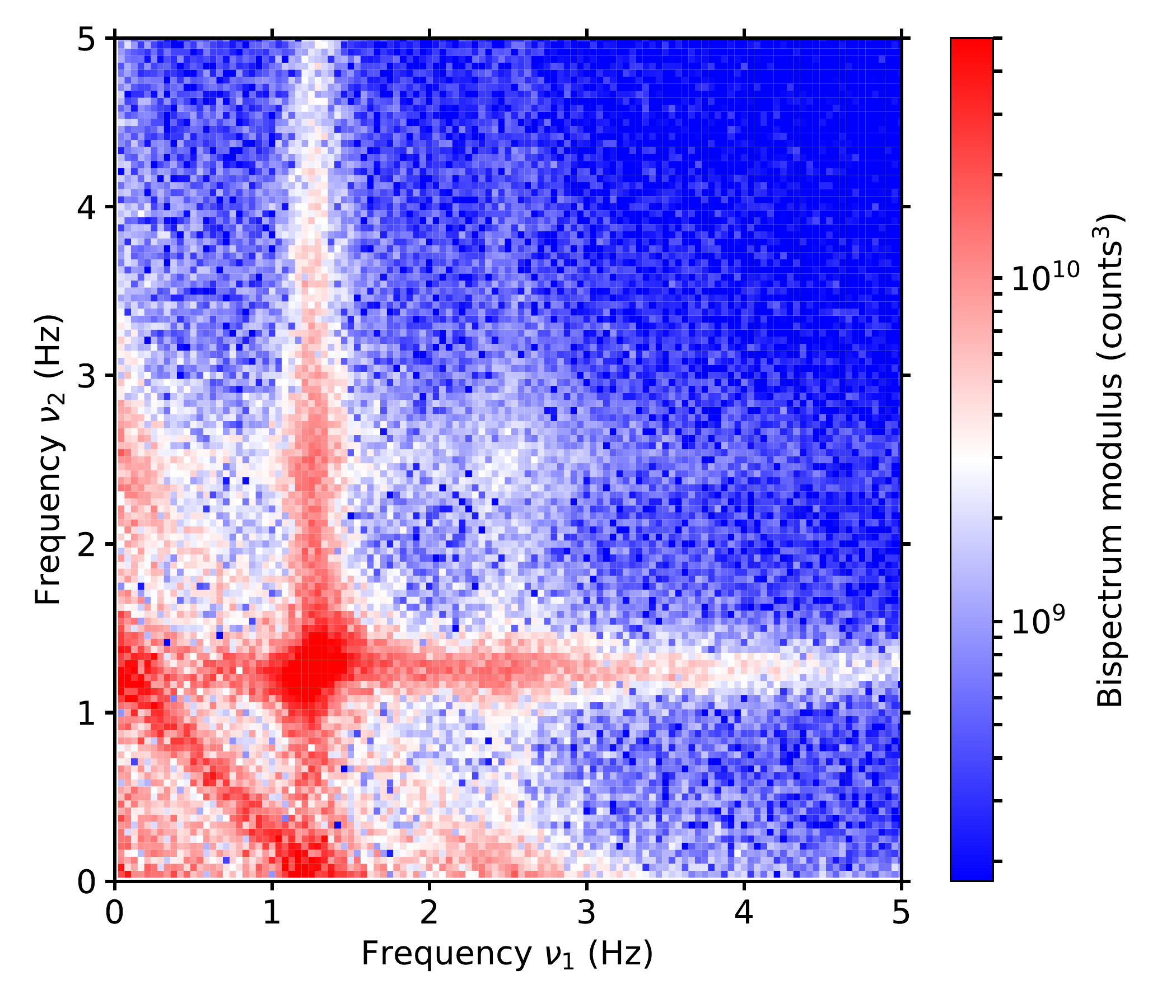}\hspace{-1mm}%
    \includegraphics[width=0.33\textwidth]{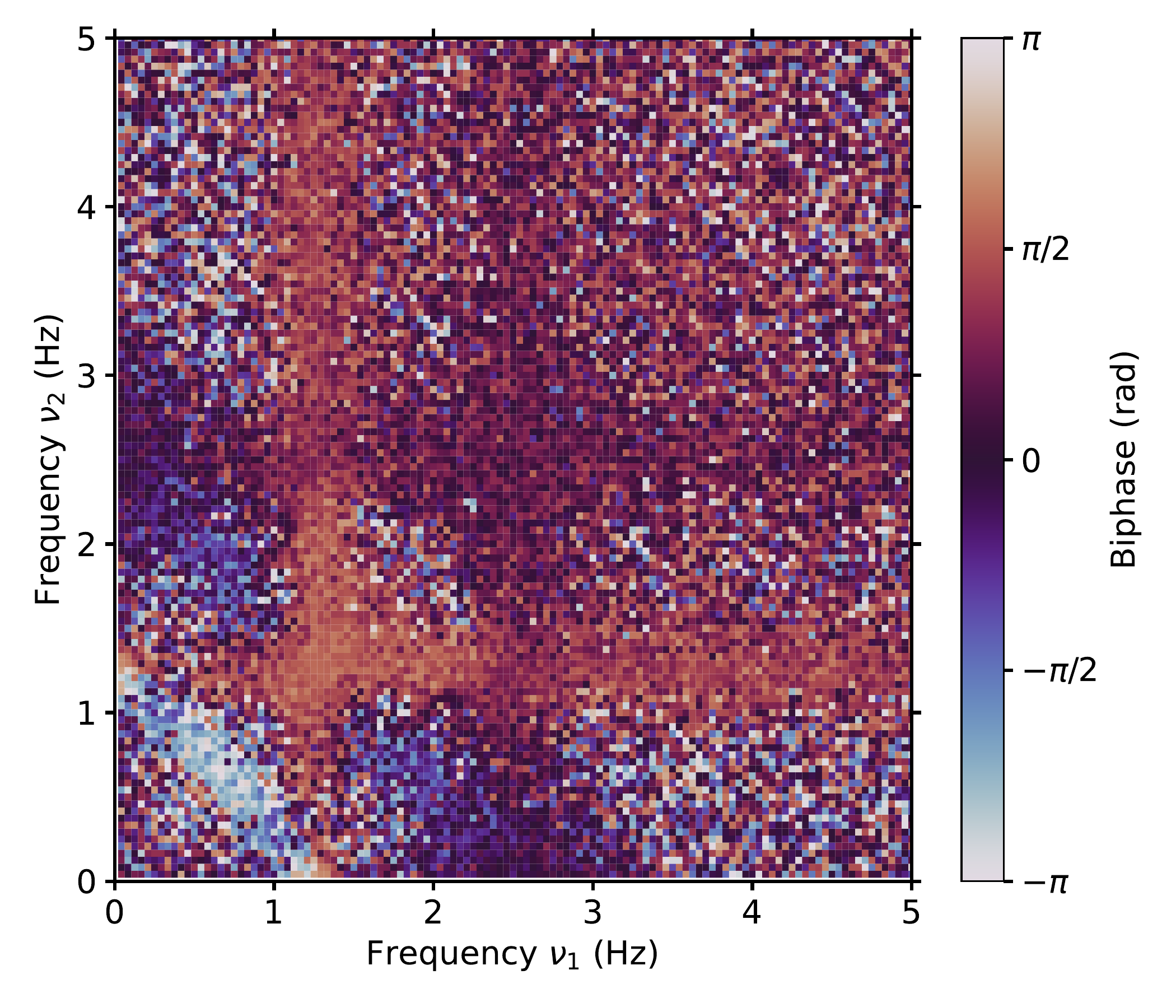}
    \vspace{-2mm}
    \caption{Bicoherence (left panel), bispectral modulus (middle panel) and biphase (right panel) maps of a NICER observation (ObsID 6203980109) of Swift~J1727.8$-$1613, with a QPO at \(1.25\) Hz.}
    \label{fig:bic_bis_bip_J1727_1.25Hz}
\end{figure*}

\begin{figure}
    \centering
    \includegraphics[width=0.33\textwidth]{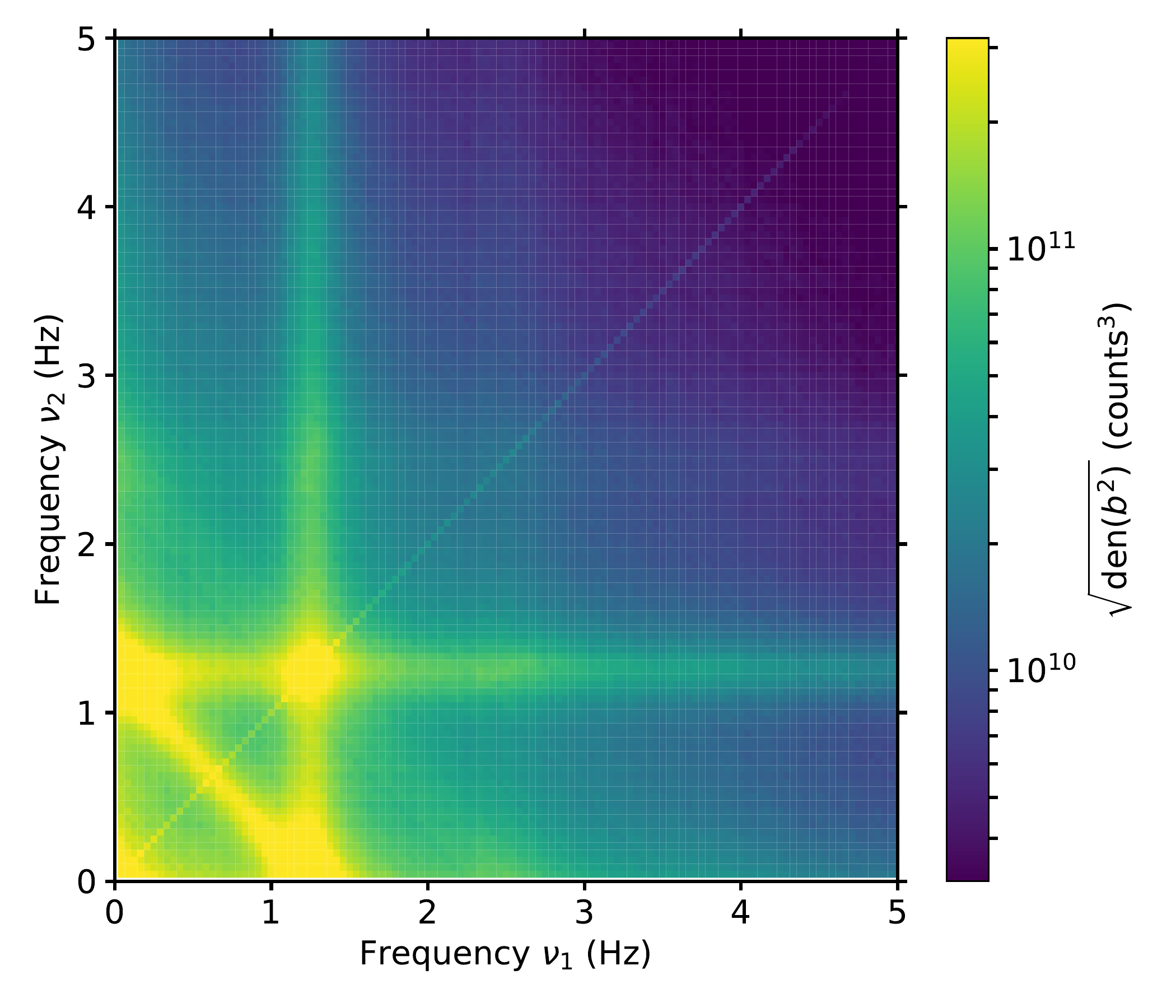}
    \vspace{-2mm}
    \caption{The square root of the bicoherence denominator of a NICER observation (ObsID 6203980109) of Swift~J1727.8$-$1613, with a QPO at \(1.25\) Hz.}
    \label{fig:bic_norm_J1727_1.25Hz}
\end{figure}

In Fig.~\ref{fig:bic_bis_bip_J1727_1.25Hz}, we show the bicoherence (left panel), bispectral modulus (middle panel), and biphase (right panel) maps of a NICER observation of Swift~J1727.8$-$1613 with a QPO at 1.25~Hz (ObsID 6203980109). This observation has \(\sim12.3\)~ks effective exposure, from which 513 segments of 24~s are used to calculate the bispectrum. 
As described above, we compute the bispectrum directly from the light curves in units of counts, without applying any
normalization or rebinning.
The bicoherence map shows a bright feature at the QPO position and horizontal and vertical ridges along the fundamental QPO and second harmonic frequencies, consistent with the so-called ``cross'' pattern reported in GRS~1915$+$105~\citep{2022MNRAS.514.1720A}. 
The bispectral-modulus map contains the same ridges but also reveals the corresponding fundamental QPO and second-harmonic hypotenuse lines, producing the so-called ``web'' pattern reported in GRS~1915$+$105~\citep{2022MNRAS.514.1720A}.
To understand this difference, Fig.~\ref{fig:bic_norm_J1727_1.25Hz} shows the Fourier-power triplet, defined as the square root of the bicoherence denominator, used to normalize \(|B|\) to obtain the bicoherence in the left panel of Fig.~\ref{fig:bic_bis_bip_J1727_1.25Hz}. The own web-like structure of the denominator indicates that a large normalization term can suppress weak coupling features in the bicoherence. Conversely, because the unnormalized bispectral modulus is weighted by Fourier amplitude, it can accentuate structures with large amplitudes and may make weak or apparent coupling features look prominent.
In the right panel of Fig.~\ref{fig:bic_bis_bip_J1727_1.25Hz}, the biphase map shows a significant ``web'' pattern. 
Importantly, the coherent biphase signal is detected only when the structure is connected with the fundamental QPO or its second harmonic, namely along the ridges of the fundamental QPO and the second harmonic and along their corresponding hypotenuse lines. No comparable coherent signal is detected away from these QPO-related structures, suggesting that the detected phase coupling is associated with the fundamental QPO and its harmonic. 

The biphase along the fundamental-QPO ridge is around $\pi/2$ rad, whereas that along the fundamental-QPO hypotenuse lies approximately between $-\pi/2$ and $-\pi$ rad. Thus, the fundamental-QPO ridge and hypotenuse appear to show opposite biphase signs, consistent with the phase relation derived above (Eq.~\ref{eq:ridge_diagonal_phase_relation}). A similar behaviour is seen for the second harmonic: the ridge has a biphase of approximately $\pi/4$ rad, while the hypotenuse has a biphase of approximately $-\pi/4$ rad (Eq.~\ref{eq:harmonic_phase_relation}). Here we use the biphase only as a qualitative diagnostic of the predicted phase relation; a quantitative analysis of its frequency dependence and statistical properties will be presented in a separate paper.

We also note distinct biphase behaviour at the intersections of different bispectral structures. This occurs where the fundamental-QPO and second-harmonic ridges overlap, as well as where a ridge intersects a hypotenuse. At these intersections, the overlapping bispectral components contribute simultaneously to the measured bispectrum through their complex sum.

\begin{figure*}
    \centering
    \includegraphics[width=0.33\textwidth]{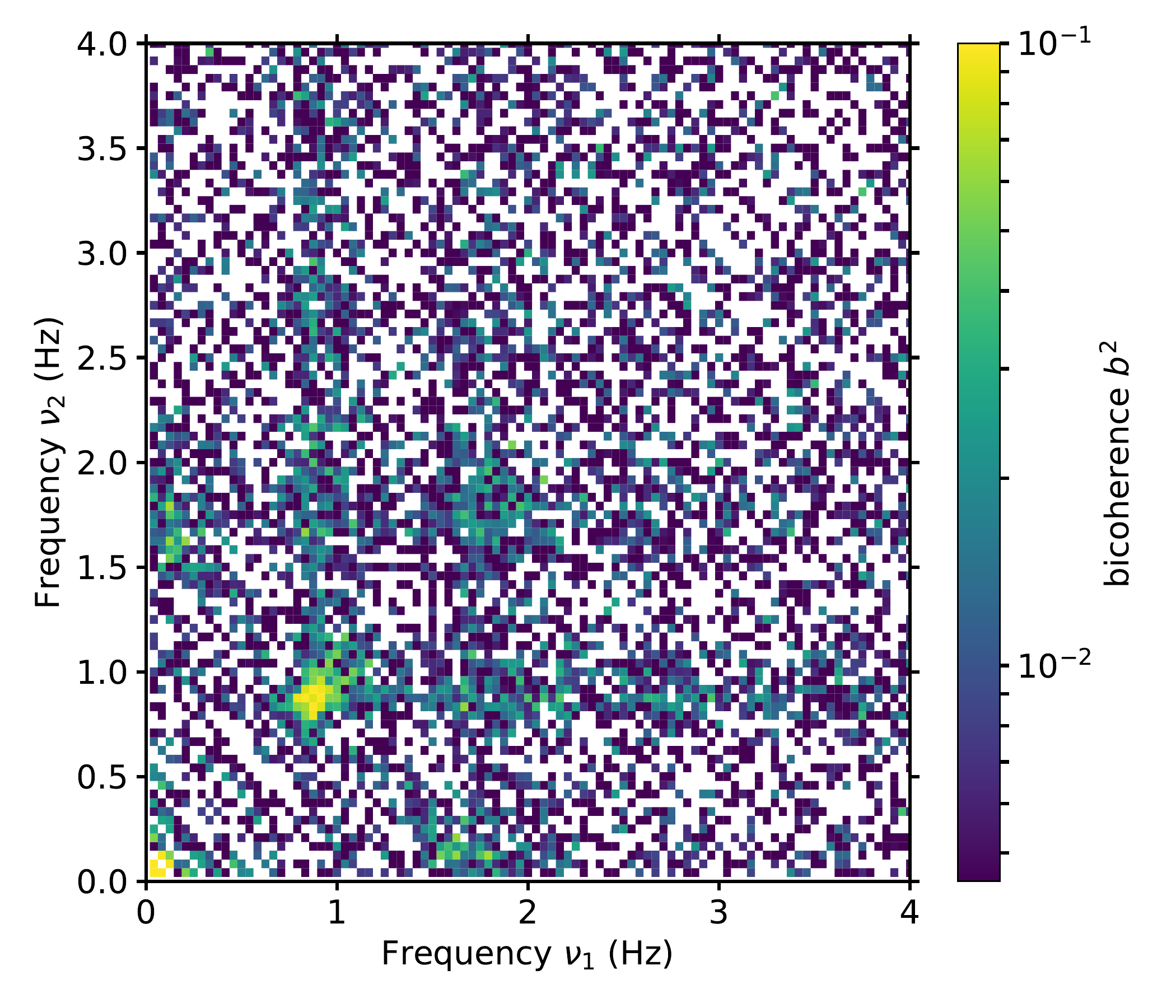}\hspace{-1mm}%
    \includegraphics[width=0.33\textwidth]{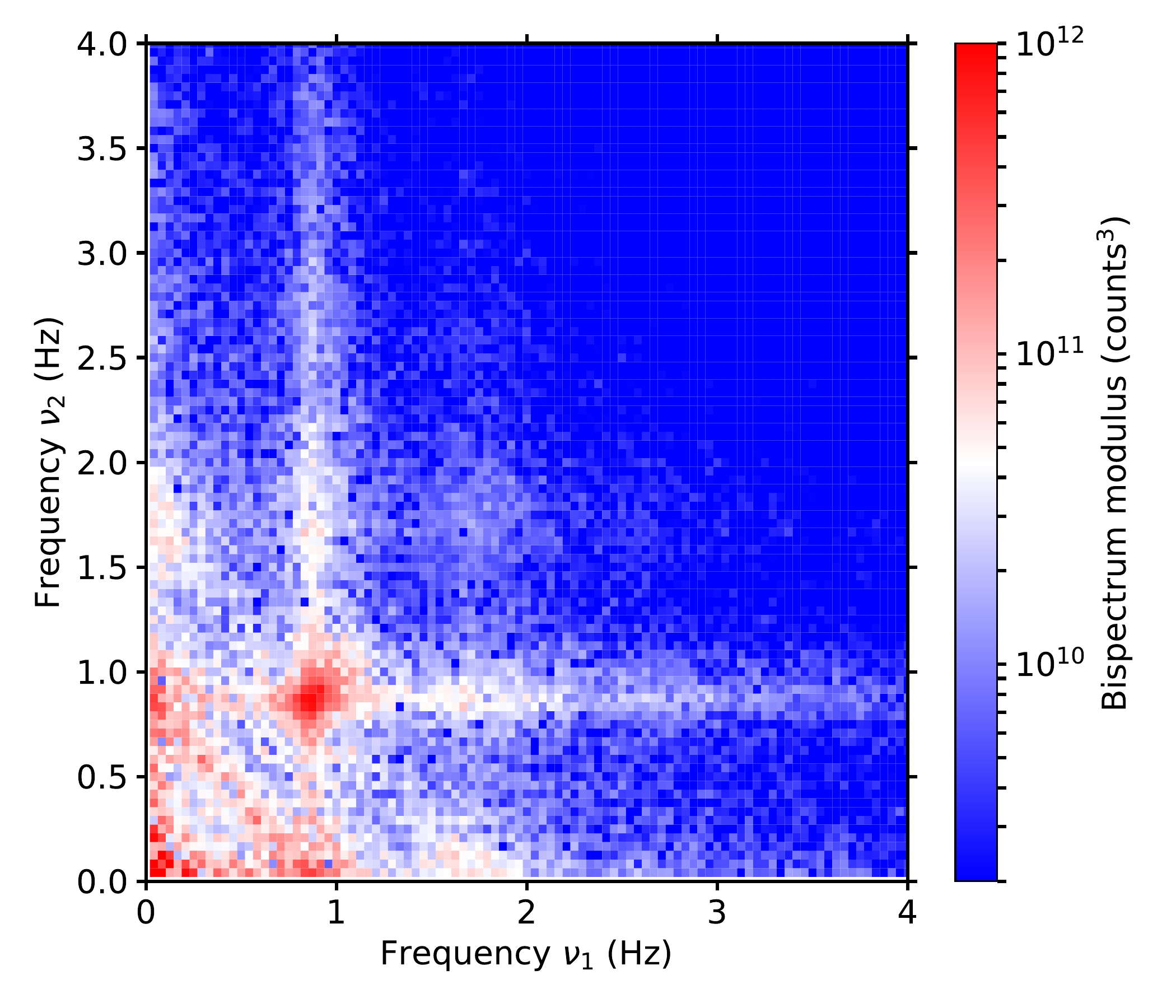}\hspace{-1mm}%
    \includegraphics[width=0.33\textwidth]{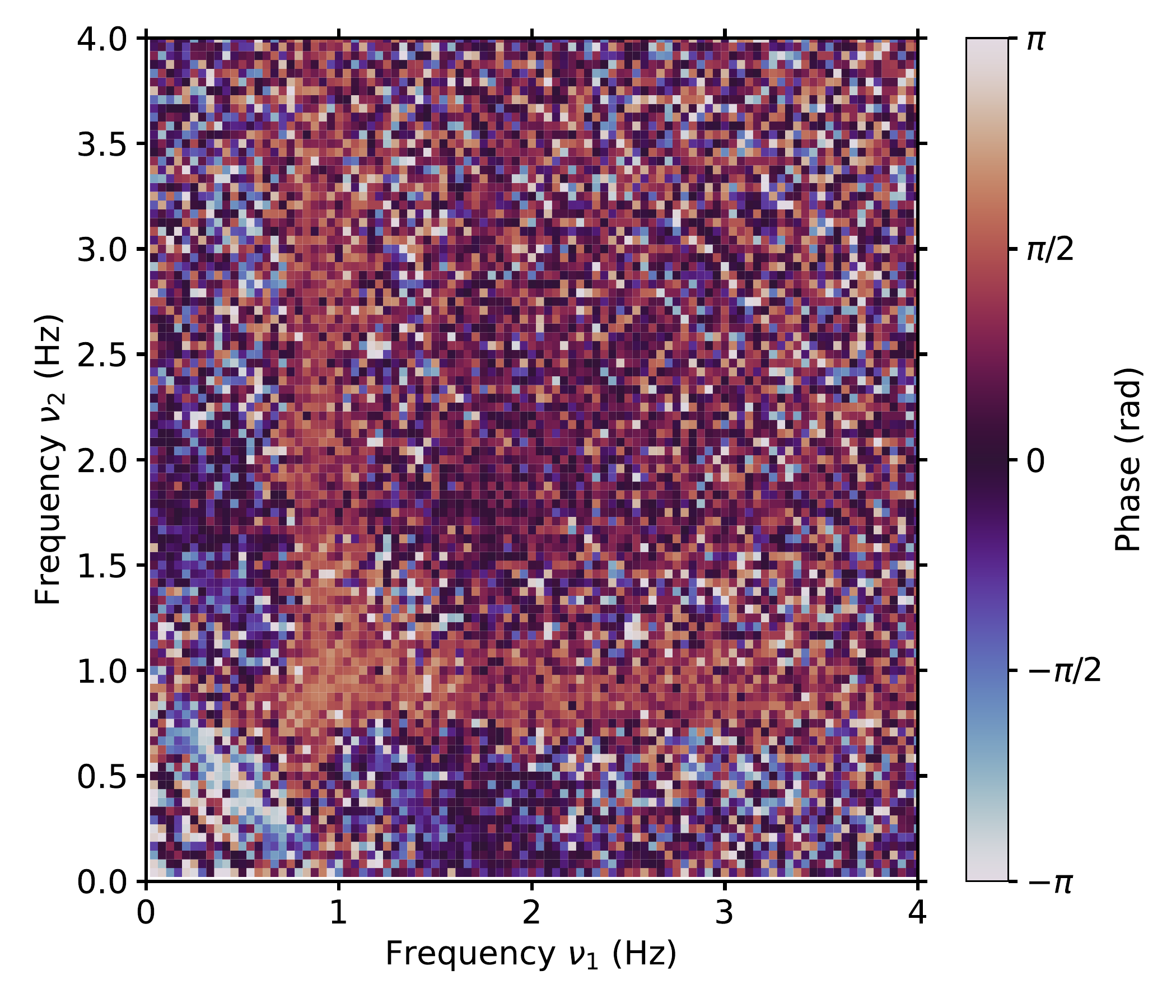}
    \vspace{-2mm}
    \caption{Same as Fig.~\ref{fig:bic_bis_bip_J1727_1.25Hz} for a NICER observation (ObsIDs 6203980105 and 6703010101) of Swift~J1727.8$-$1613, with a QPO at \(0.88\) Hz.}
    \label{fig:bic_bis_bip_J1727_0.875Hz}
\end{figure*}

To test whether the same coupling pattern persists at a lower QPO frequency, we examine a second NICER observation of Swift~J1727.8$-$1613 with a QPO at 0.88~Hz (ObsIDs 6203980105 and 6703010101). Its effective exposure of \(\sim\)5.3~ks provides 221 segments of 24~s for the bispectral calculation. Despite the shorter exposure, the bicoherence map (left panel of Fig.~\ref{fig:bic_bis_bip_J1727_0.875Hz}) again shows a localized enhancement at the QPO position and weaker ridges connected to the fundamental QPO and the second harmonic. The corresponding bispectral-modulus map (middle panel of Fig.~\ref{fig:bic_bis_bip_J1727_0.875Hz}) displays a clearer web-like pattern, with enhanced signal along the fundamental QPO ridge and hypotenuse, as expected because the bispectrum is weighted by the Fourier amplitudes. The biphase (right panel of Fig.~\ref{fig:bic_bis_bip_J1727_0.875Hz}) remains coherent mainly on these QPO-related structures and becomes scattered elsewhere, suggesting again that the detected phase coupling is tied to the fundamental QPO and its harmonic. 
Moreover, the approximate opposite-sign relation between the ridge and hypotenuse biphases is again apparent for both the fundamental QPO and the second harmonic, consistent with the behavior found in the observation with the QPO at 1.25~Hz and with the qualitative prediction of the modulation model of biphases with opposite-signs across the ridge and corresponding hypotenuse.

\begin{figure*}
    \centering
    \includegraphics[width=0.33\textwidth]{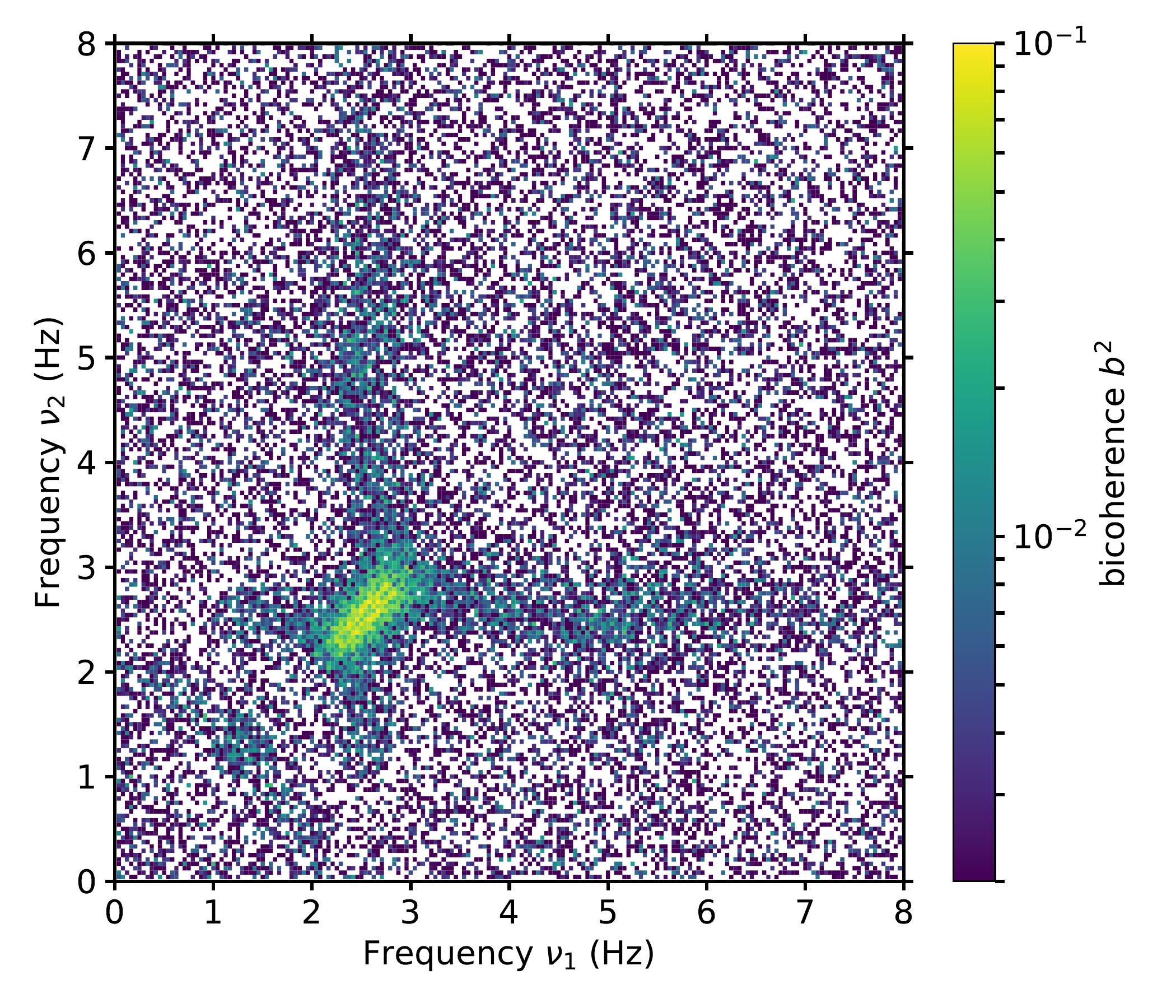}\hspace{-1mm}%
    \includegraphics[width=0.33\textwidth]{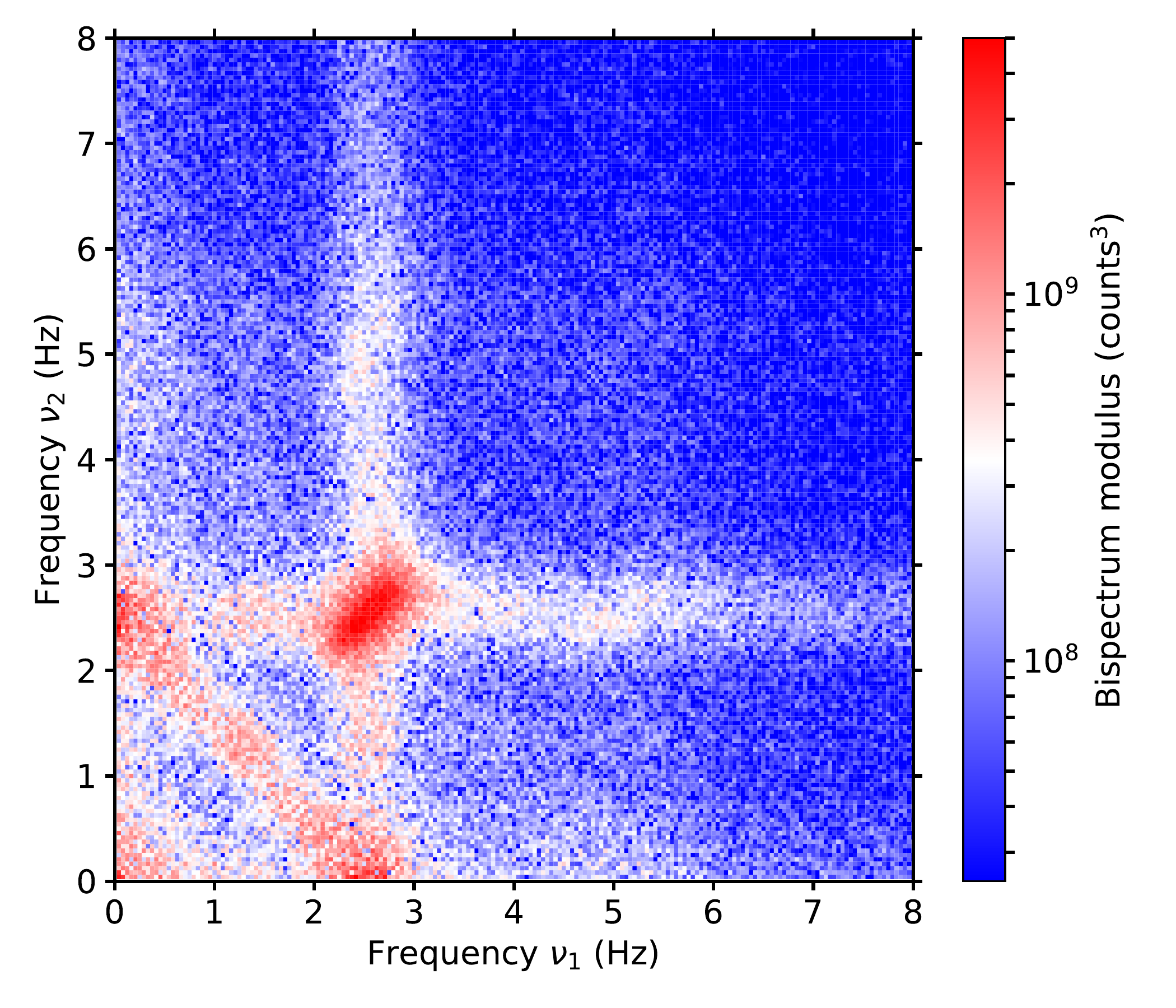}\hspace{-1mm}%
    \includegraphics[width=0.33\textwidth]{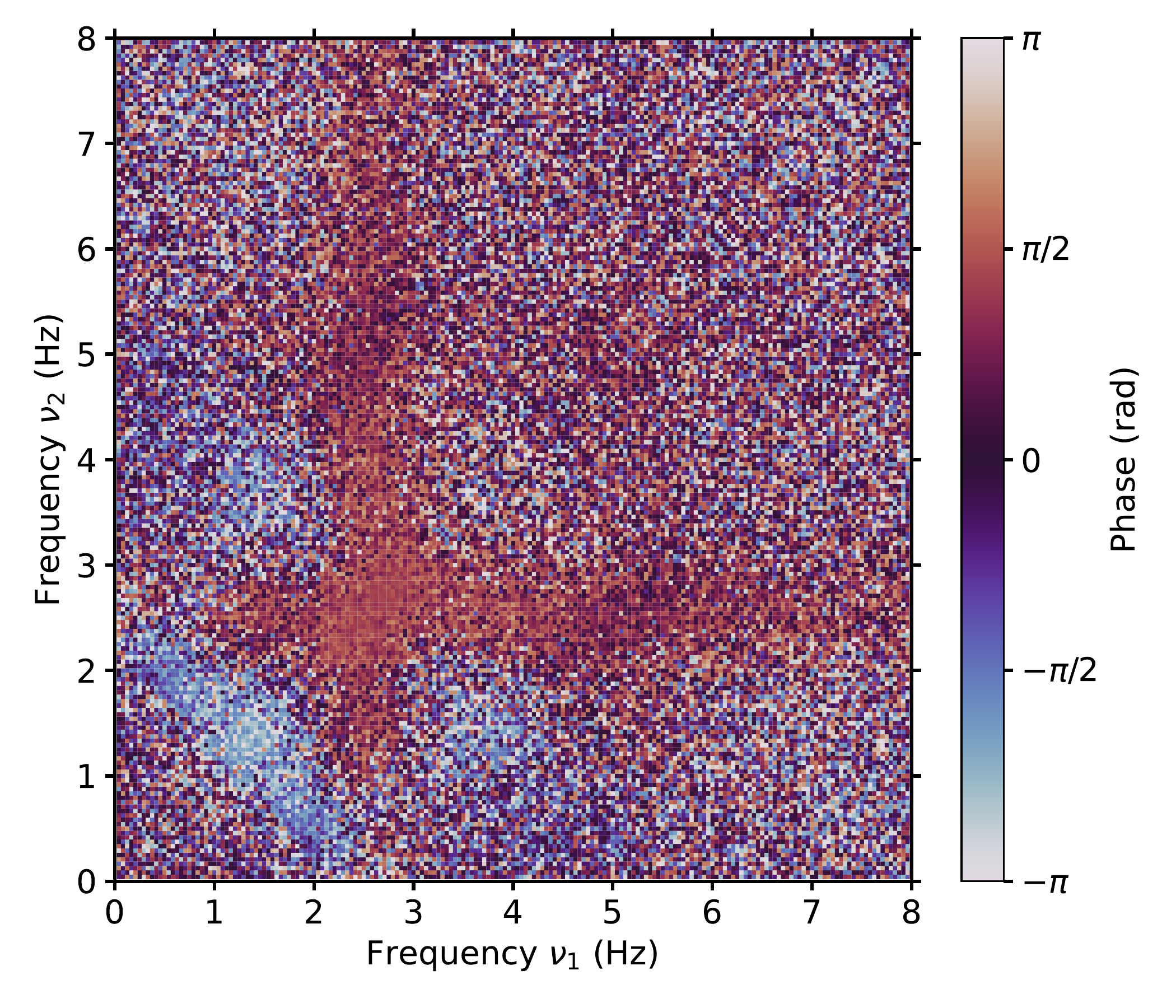}
    \vspace{-2mm}
    \caption{Same as Fig.~\ref{fig:bic_bis_bip_J1727_1.25Hz} for a NICER observation (ObsIDs 6703010107 and 6750010202) of Swift~J1727.8$-$1613, with a QPO at \(2.54\) Hz.}
    \label{fig:bic_bis_bip_J1727_2.67Hz}
\end{figure*}

For the higher-frequency QPO (2.54~Hz) NICER observation of Swift~J1727.8$-$1613 (ObsIDs 6703010107 and 6750010202) shown in Fig.~\ref{fig:bic_bis_bip_J1727_2.67Hz}, the effective exposure is \(\sim\)10.5~ks, yielding 438 segments of 24~s for the bispectral calculation. The bicoherence, bispectral modulus, and biphase maps show the same overall behavior as in the lower-frequency observations: coherent structures remain associated with the fundamental QPO and its harmonic, and the ridge--hypotenuse pattern is still present. However, the ridge and hypotenuse structures associated with the second harmonic are much weaker in this observation.
We note that the bicoherence exhibits an elongated feature around the QPO frequency pair, which is caused by the QPO frequency drifting between $\sim$2.0 and $\sim$3.0 Hz during this observation (see Appendix~\ref{lab:Dynamical power spectrum}).

\subsubsection{MAXI~J1820$+$070}
\label{subsubsec:MAXI~J1820$+$070}

\begin{figure*}
    \centering
    \includegraphics[width=0.33\textwidth]{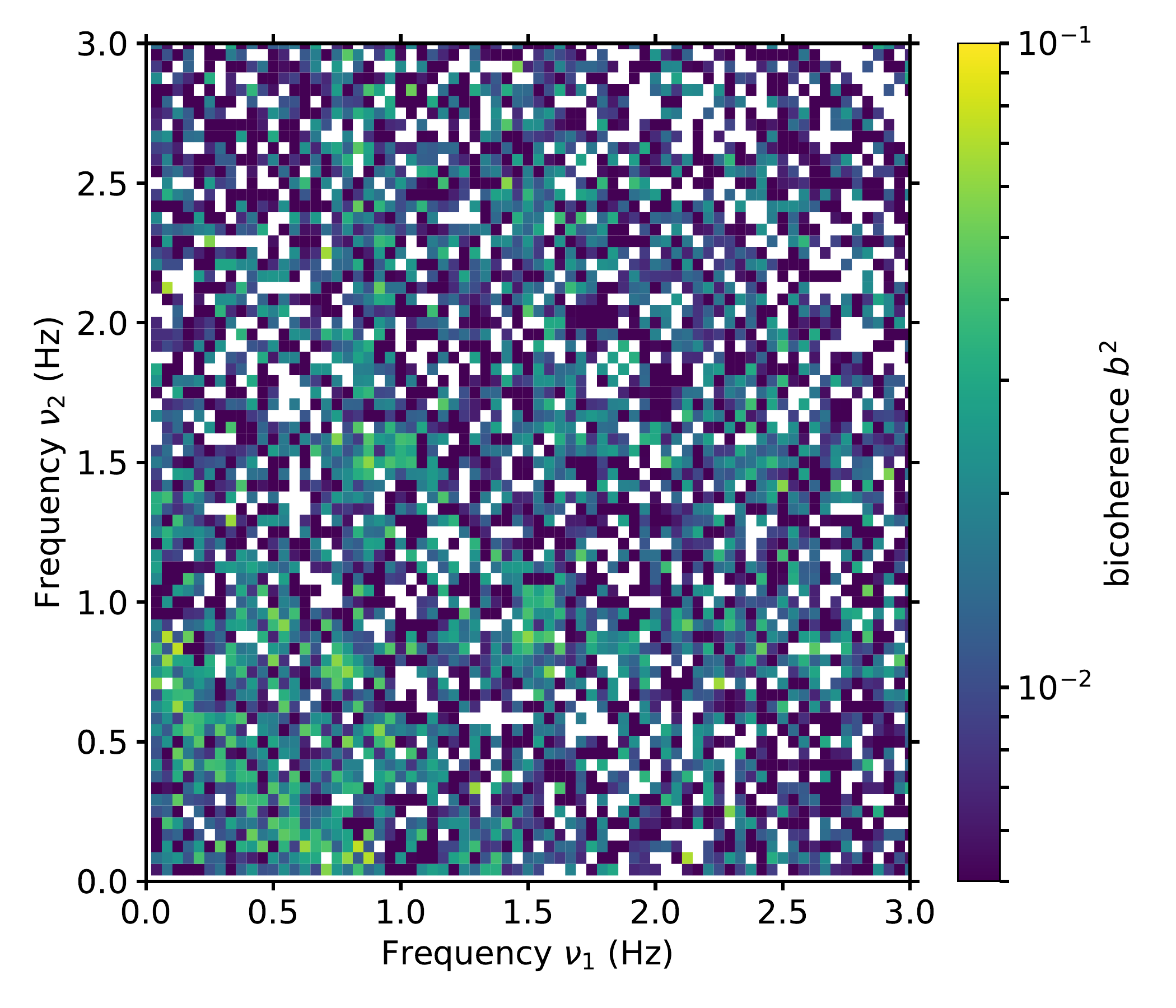}\hspace{-1mm}%
    \includegraphics[width=0.33\textwidth]{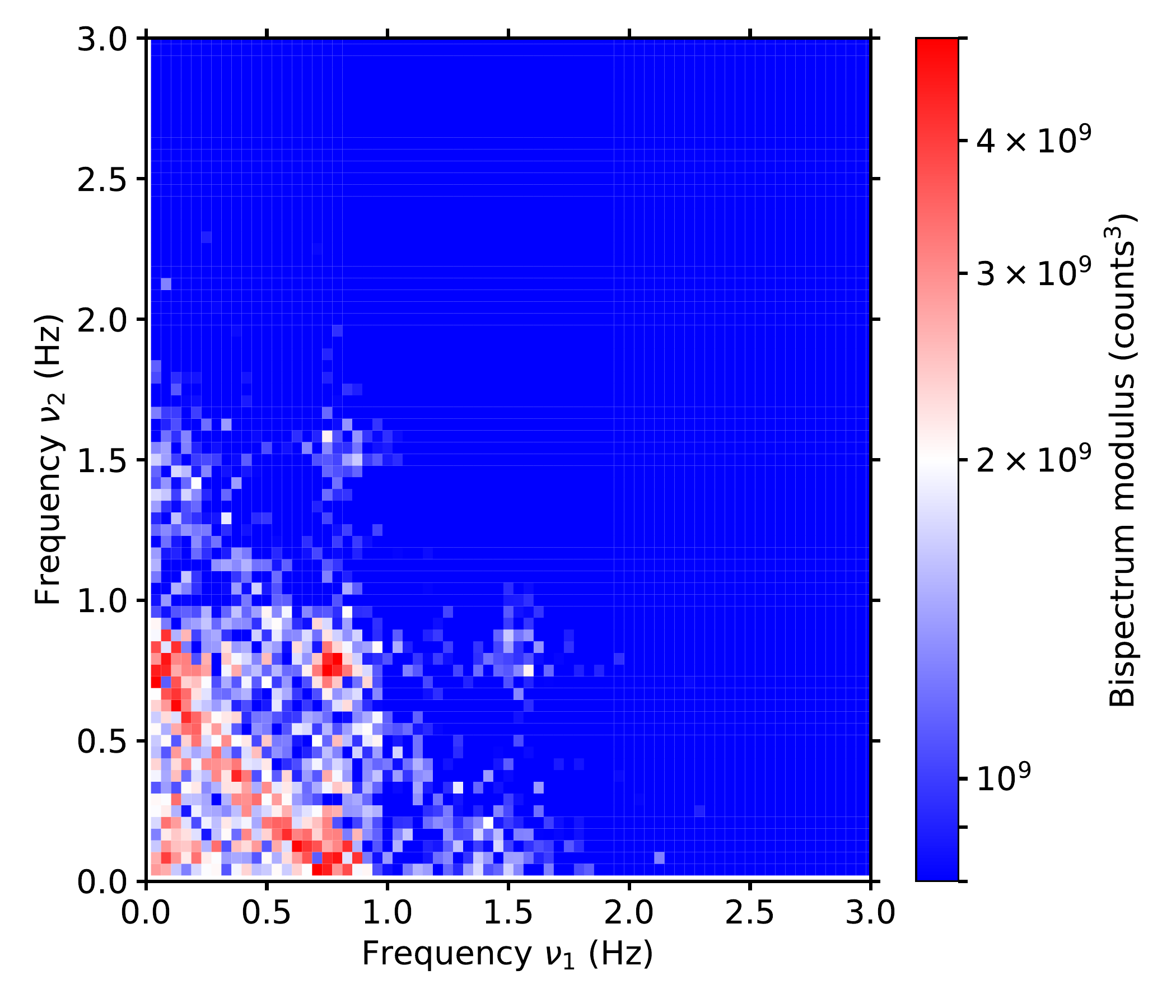}\hspace{-1mm}%
    \includegraphics[width=0.33\textwidth]{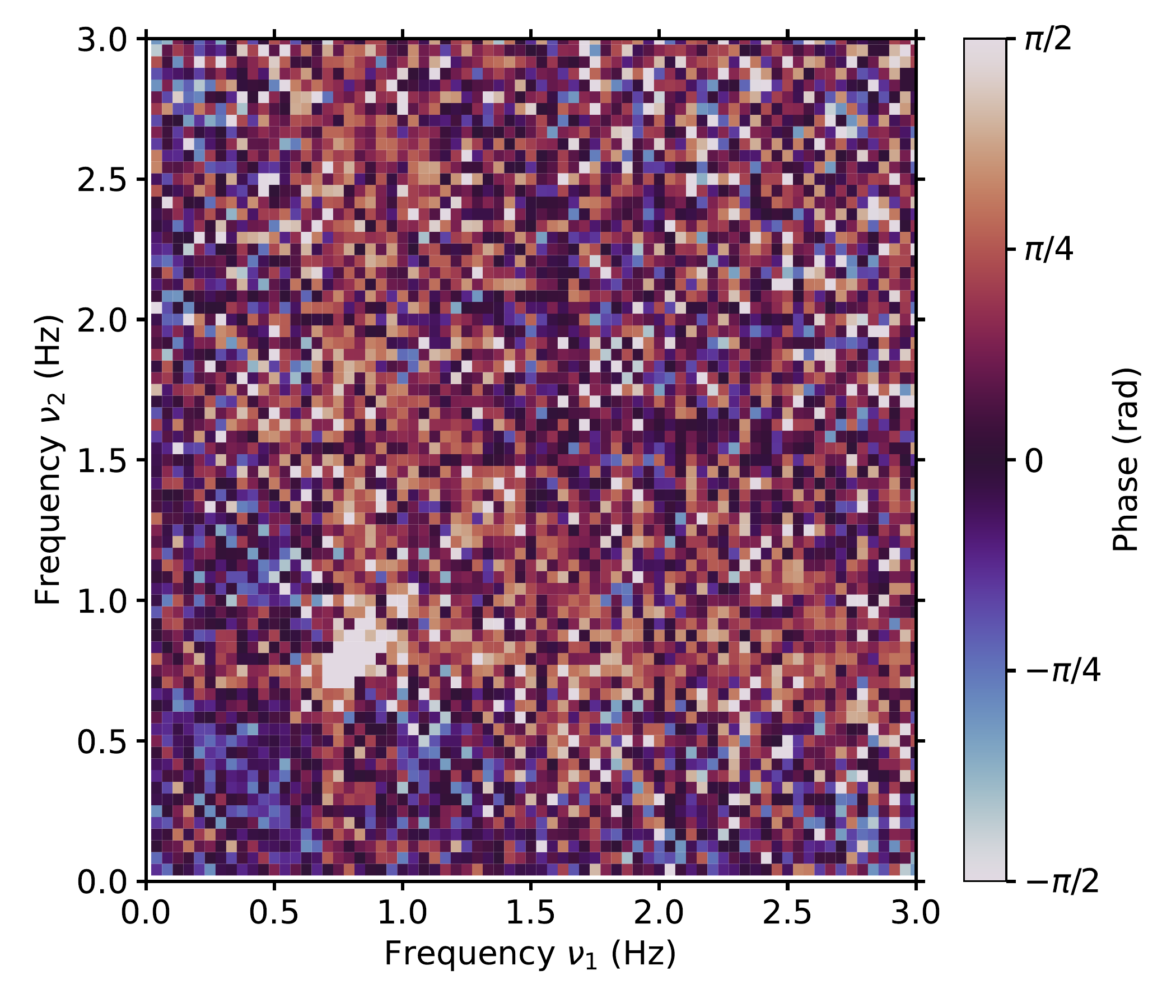}
    \vspace{-2mm}
    \caption{Same as Fig.~\ref{fig:bic_bis_bip_J1727_1.25Hz} for a NICER observation (ObsID 1200120194) of MAXI~J1820$+$070, with a QPO at \(0.75\) Hz.}
    \label{fig:bic_bis_bip_J1820_0.75Hz}
\end{figure*}

In Fig.~\ref{fig:bic_bis_bip_J1820_0.75Hz}, we show the bicoherence (left panel), bispectral modulus (middle panel), and biphase (right panel) maps of the NICER observation, ObsID 1200120194, of MAXI~J1820$+$070 with a QPO at 0.75~Hz. 
The effective exposure is \(\sim\)5.3~ks, yielding 217 segments of 24~s for the bispectral calculation. 
We select this observation because it provides the most suitable balance for bispectral analysis: at higher QPO frequencies, the QPO frequency evolves too rapidly during an observation, whereas at lower QPO frequencies, the signal-to-noise ratio is reduced by stronger broadband variability.
Unlike the observations of Swift~J1727.8$-$1613, the bicoherence map does not show a clear so-called ``cross'', ``web'' or ``hypotenuse''-like pattern of the type reported for GRS~1915$+$105~\citep{2022MNRAS.514.1720A}. The bispectral-modulus map shows a significant hypotenuse feature associated with the fundamental QPO, together with a weaker fundamental-QPO ridge and a weak hypotenuse feature associated with the second harmonic. The biphase map, importantly, reveals a negative biphase along the fundamental-QPO hypotenuse and a positive biphase along the corresponding ridge. These opposite-sign phase structures form a recognizable ``web''-like pattern, qualitatively consistent with the prediction of the multiplicative-modulation model.

\subsection{Fundamental QPO ridge}
\label{subsec:fundamental-ridge}

\begin{figure}
    \centering
    \vspace{-2mm}    
    \includegraphics[width=0.85\columnwidth]{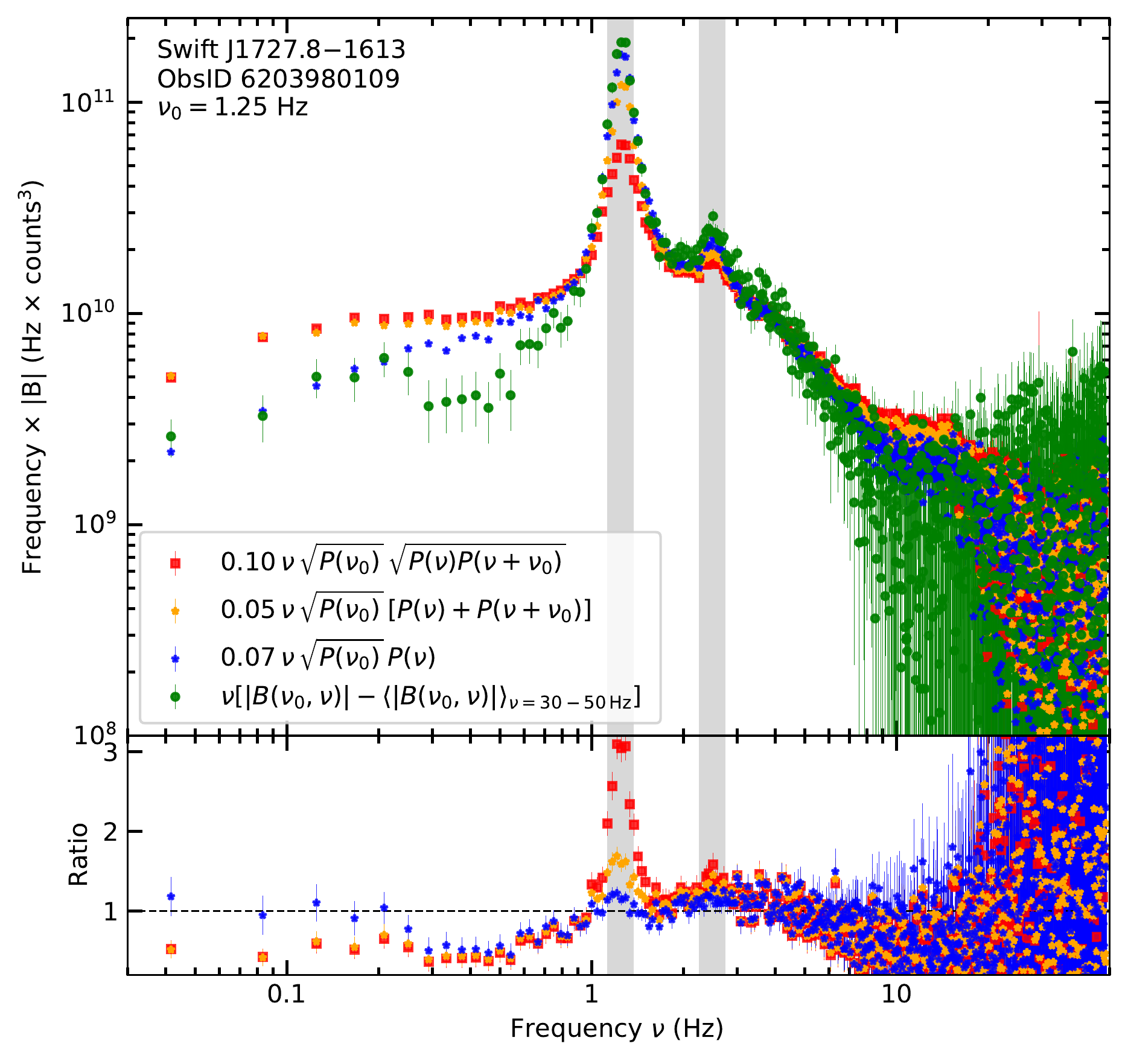}
    \includegraphics[width=0.85\columnwidth]{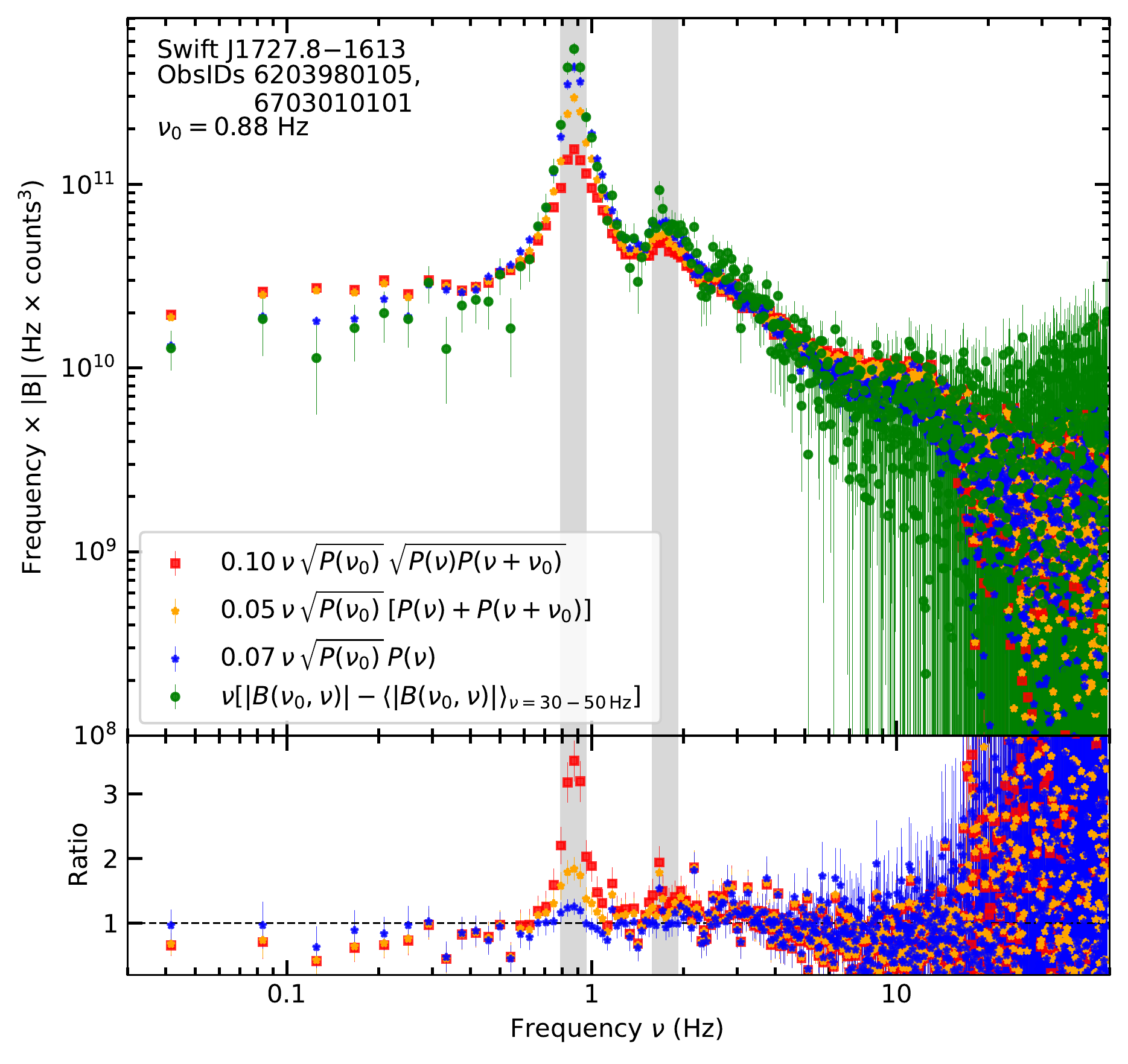}
    \includegraphics[width=0.85\columnwidth]{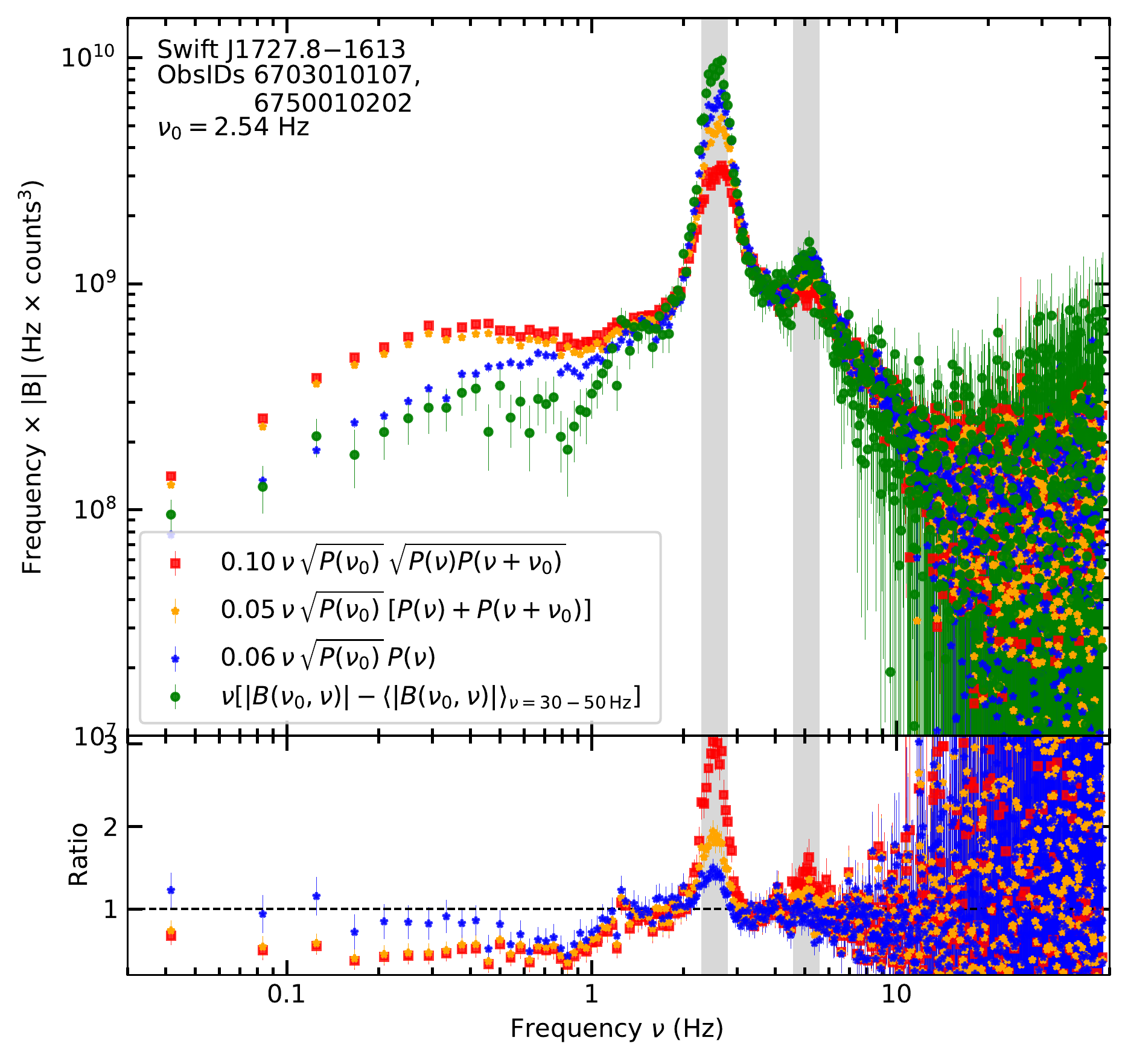}
    \vspace{-2mm}
    \caption{
    Modulus of the bispectra along the fundamental QPO ridge for the three NICER observations of Swift~J1727.8$-$1613. The green points show the bispectral modulus, \(|B(\nu_0,\nu)|\), after subtracting the average value over the 30--50~Hz frequency range, \(\langle |B(\nu_0,\nu)| \rangle_{\rm \nu=30-50~Hz}\). The red points show the scaled Fourier-power triplet, \(\sqrt{P(\nu_0)}\sqrt{P(\nu)P(\nu+\nu_0)}\). The orange points show the scaled modulation-model prediction, \(\sqrt{P(\nu_0)}[P(\nu)+P(\nu+\nu_0)]\) (see Eq.~\ref{eq:qpo_ridge_mag}). The blue points show the scaled power spectrum, \(\sqrt{P(\nu_0)}P(\nu)\) (see Eq.~\ref{eq:qpo_ridge_mag_ltoh}).
    In each plot the lower panels show the ratios between the bispectral modulus and the three quantities in the upper panels.
    The gray shaded regions highlight the frequency ranges of the fundamental QPO and its second harmonic.}
    \label{fig:qpo_ridge_J1727}
\end{figure}

In Fig.~\ref{fig:qpo_ridge_J1727}, we show the modulus of the bispectrum (green) along the fundamental QPO ridge, \(|B(\nu_0,\nu)|\), for the NICER observations of Swift~J1727.8$-$1613 presented in Section~\ref{subsubsec:Swift~J1727.8$-$1613}. 
The bispectral modulus is extracted from the two-dimensional bispectral-modulus map by averaging across the fundamental QPO ridge over a frequency width set by \(W_1=0.2\), without binning along the ridge.
Since the measured bispectral modulus has a positive noise contribution, we subtract its average value between 30 and 50 Hz, where the intrinsic bispectral signal is expected to be negligible. This provides an approximate correction for the noise contribution, rather than an unbiased estimate of the intrinsic bispectral modulus.
In all three observations, the ridge modulus peaks at the fundamental QPO and second-harmonic frequencies; above \(\sim10\) Hz, the measurements become increasingly noisy and scattered. The red points show the scaled Fourier-power triplet, \(\sqrt{P(\nu_0)}\sqrt{P(\nu)P(\nu+\nu_0)}\). The orange points show the scaled modulation-model prediction, \(\sqrt{P(\nu_0)}\left[P(\nu)+P(\nu+\nu_0)\right]\), from Eq.~\ref{eq:qpo_ridge_mag}. 
Here, we approximate $a_1/2$ in Eq.~\ref{eq:qpo_ridge_mag} by $\sqrt{P(\nu_0)}$ and absorb the modulation-amplitude factor $u_1/2$ into the fitted scaling factor.
The blue points show the scaled power-spectrum term, \(\sqrt{P(\nu_0)}P(\nu)\), interpreted as the one-sided form of the modulation model in which variability is shifted from lower to higher frequencies, while the reverse shift is suppressed (see Eq.~\ref{eq:qpo_ridge_mag_ltoh}). The lower panels show the ratios between the bispectral modulus and the quantities used for comparsion.

All three quantities broadly reproduce the broad continuum trend of the bispectral modulus. At the lowest frequencies, the ratios for the Fourier-power triplet and the modulation-model prediction are below unity, indicating that these two quantities overestimate the measured bispectral modulus. Around the fundamental QPO frequency (the gray region), however, the Fourier-power triplet substantially underestimates the measured bispectral modulus, producing a pronounced peak in the corresponding ratio. The modulation-model prediction reduces this discrepancy but still underestimates the bispectral modulus around the QPO. The power-spectrum term provides the closest agreement in the QPO region, with its ratio remaining closer to unity, although a slight residual discrepancy remains.
We caution, however, that the region around the QPO frequencies includes intersections between different bispectral structures, including the fundamental-QPO ridge and the second-harmonic hypotenuse, with an additional contribution from intrinsic QPO--harmonic coupling. The isolated ridge prediction is therefore not expected to work directly in this region (see Section~\ref{subsec:intrinsic_qpo_harmonic} and Appendix~\ref{app:intersections}).

\begin{figure}
    \centering
    \includegraphics[width=0.85\columnwidth]{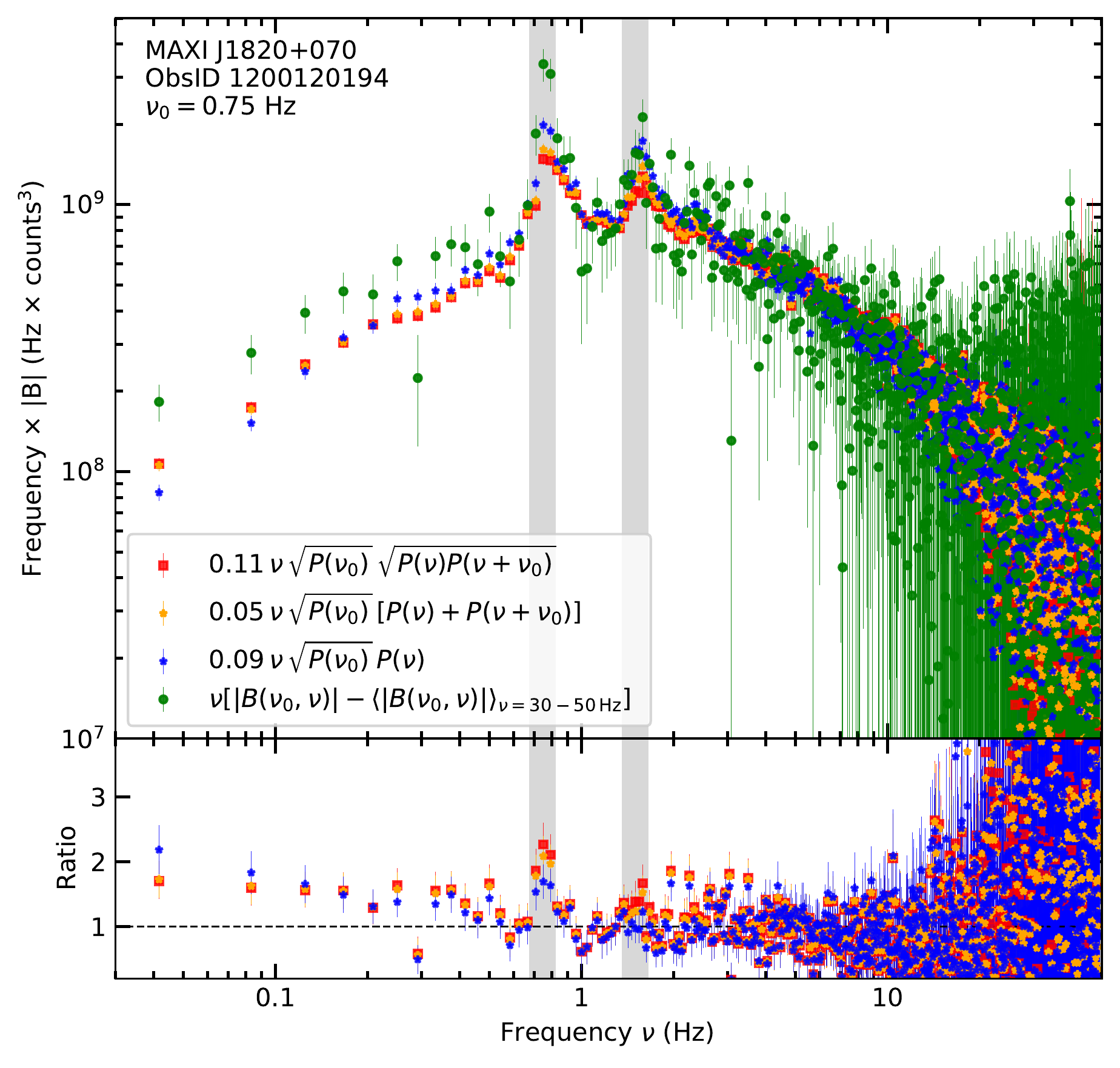}
    \vspace{-2mm}
    \caption{Bispectral modulus along the fundamental QPO ridge of MAXI~J1820$+$070. The plotted quantities and color scheme are the same as those used in Fig.~\ref{fig:qpo_ridge_J1727}.}
    \label{fig:qpo_ridge_J1820}
\end{figure}

Fig.~\ref{fig:qpo_ridge_J1820} shows the bispectral modulus along the fundamental QPO ridge of MAXI~J1820$+$070. The three quantities all broadly reproduce the overall frequency dependence of the bispectral modulus. Around the fundamental QPO frequency (the gray region), as in Swift~J1727.8$-$1613, the power-spectrum term most closely follows the measured bispectral modulus around the QPO-related peaks.

\subsection{Second-harmonic ridge}
\label{subsec:Second-harmonic ridge}

\begin{figure}
    \centering
    \vspace{-2mm}
    \includegraphics[width=0.85\columnwidth]{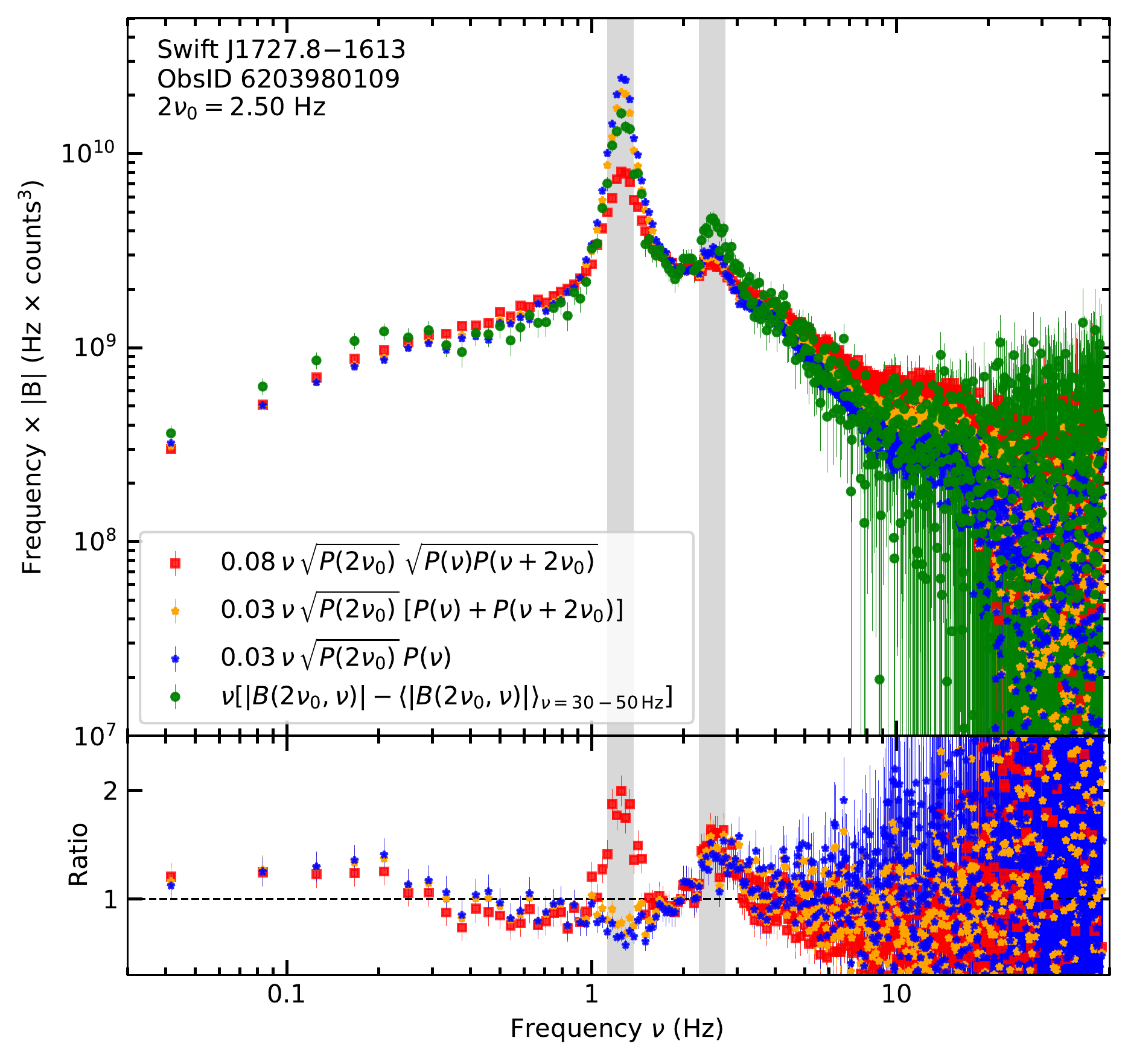}
    \includegraphics[width=0.85\columnwidth]{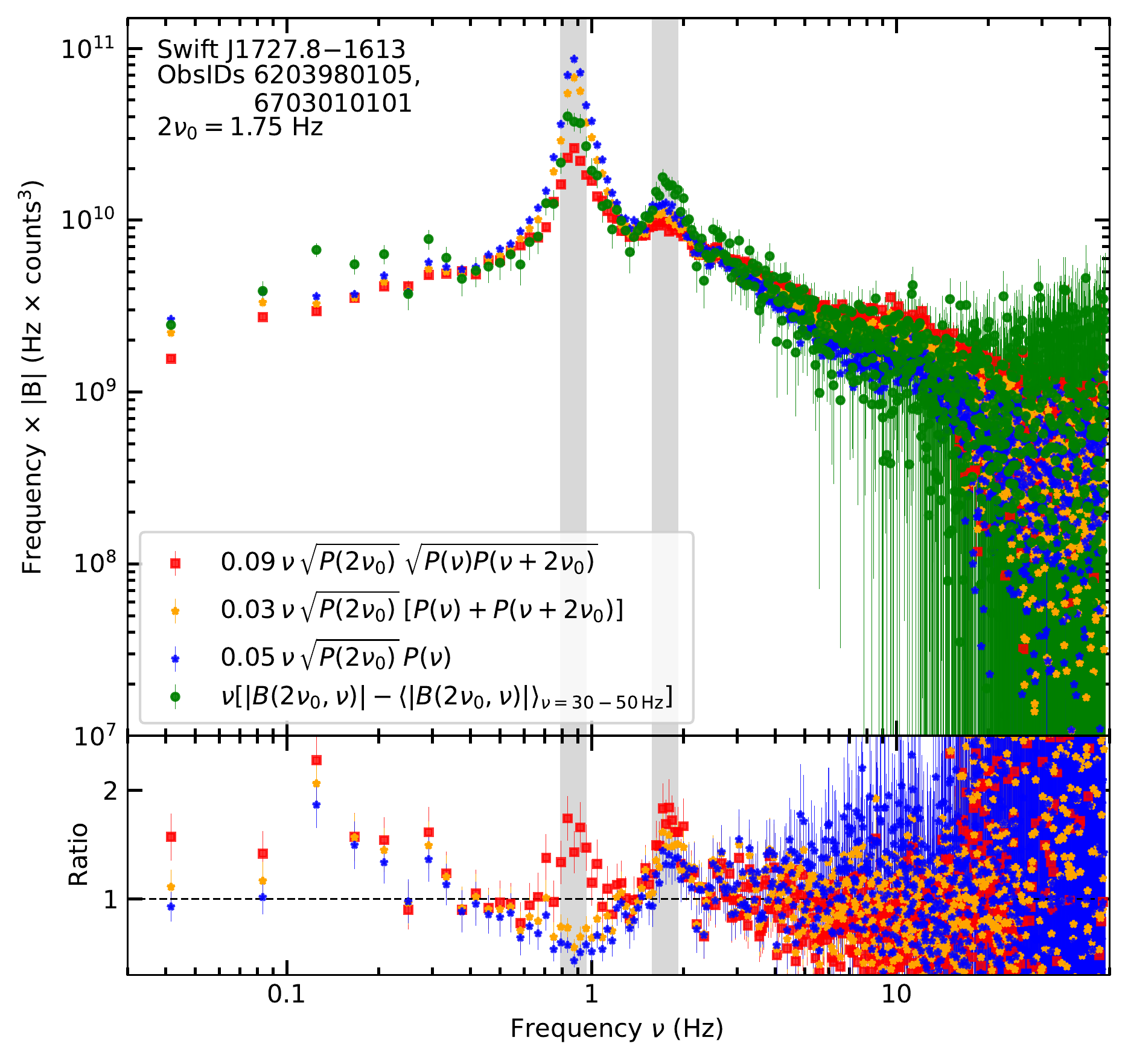}
    \includegraphics[width=0.85\columnwidth]{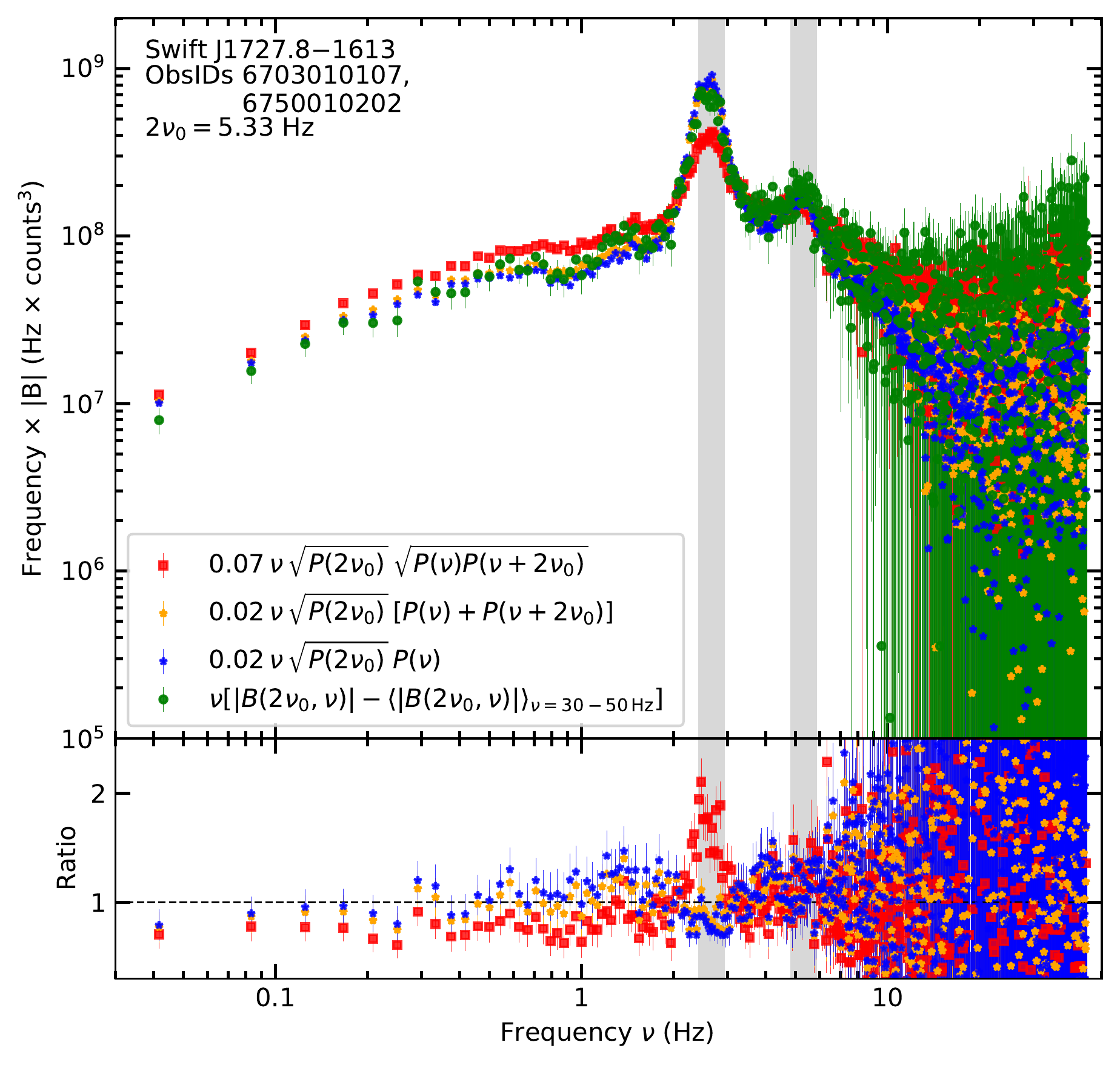}
    \vspace{-2mm}
    \caption{Modulus of the bispectra along the second harmonic ridge for the three NICER observations of Swift~J1727.8$-$1613. The green points show the bispectral modulus, \(|B(2\nu_0,\nu)|\), after subtracting the average value over the 30--50~Hz frequency range, \(\langle |B(2\nu_0,\nu)| \rangle_{\rm \nu=30-50~Hz}\). The red points show the scaled Fourier-power triplet, \(\sqrt{P(2\nu_0)}\sqrt{P(\nu)P(\nu+2\nu_0)}\). The orange points show the scaled modulation-model prediction, \(\sqrt{P(2\nu_0)}[P(\nu)+P(\nu+2\nu_0)]\) (see Eq.~\ref{eq:harmonic_ridge_mag}). The blue points show the scaled power spectrum, \(\sqrt{P(2\nu_0)}P(\nu)\) (see Eq.~\ref{eq:harmonic_ridge_mag_ltoh}).
    In each plot the lower panels show the ratios between the bispectral modulus and the three quantities in the upper panels.
    The gray shaded regions highlight the frequency ranges of the fundamental QPO and its second harmonic.}
    \label{fig:harmonic_ridge_J1727}
\end{figure}

Fig.~\ref{fig:harmonic_ridge_J1727} shows the bispectral modulus (green) along the second-harmonic QPO ridge, \(|B(2\nu_0,\nu)|\), for the same three NICER observations of Swift~J1727.8$-$1613. In all three observations, the bispectral modulus exhibits a prominent peak near the fundamental QPO frequency and a secondary enhancement around the second harmonic. At frequencies above approximately \(10~{\rm Hz}\), the measurements become increasingly noise dominated and show substantial scatter.

For comparison, the figure also shows the scaled Fourier-power triplet, \(\sqrt{P(2\nu_0)}\sqrt{P(\nu)P(\nu+2\nu_0)}\), the scaled modulation-model prediction, \(\sqrt{P(2\nu_0)}[P(\nu)+P(\nu+2\nu_0)]\) (Eq.~\ref{eq:harmonic_ridge_mag}), and the scaled power-spectrum term, \(\sqrt{P(2\nu_0)}P(\nu)\) (Eq.~\ref{eq:harmonic_ridge_mag_ltoh}). 
For the latter terms, we approximate $a_2/2$ in Eq.~\ref{eq:qpo_ridge_mag} by $\sqrt{P(2\nu_0)}$ and absorb the modulation-amplitude factor $u_2/2$ into the fitted scaling factor.
Although all three quantities reproduce the overall broadband frequency dependence, the Fourier-power triplet shows positive residuals at the fundamental QPO and second-harmonic peaks (the gray regions), indicating that it underestimates the bispectral modulus at these frequencies.
In all three observations, the modulation-model prediction and the power-spectrum term yield nearly identical values and require very similar, or even identical, scaling factors. This occurs because, when \(P(\nu)\) dominates over \(P(\nu+2\nu_0)\), the modulation-model prediction,
\(
\sqrt{P(2\nu_0)}\left[P(\nu)+P(\nu+2\nu_0)\right],
\)
reduces approximately to the power-spectrum term,
\(
\sqrt{P(2\nu_0)}P(\nu).
\)
Both the modulation-model prediction and the power-spectrum term show $<1$ ratio at the fundamental QPO, but $>1$ ratio at the second harmonic.
These frequencies coincide with regions where multiple bispectral structures overlap, and the isolated comparison is therefore not expected to fully describe the measured bispectral modulus, as discussed in Appendix~\ref{app:intersections}.

\begin{figure}
    \centering
    \includegraphics[width=0.85\columnwidth]{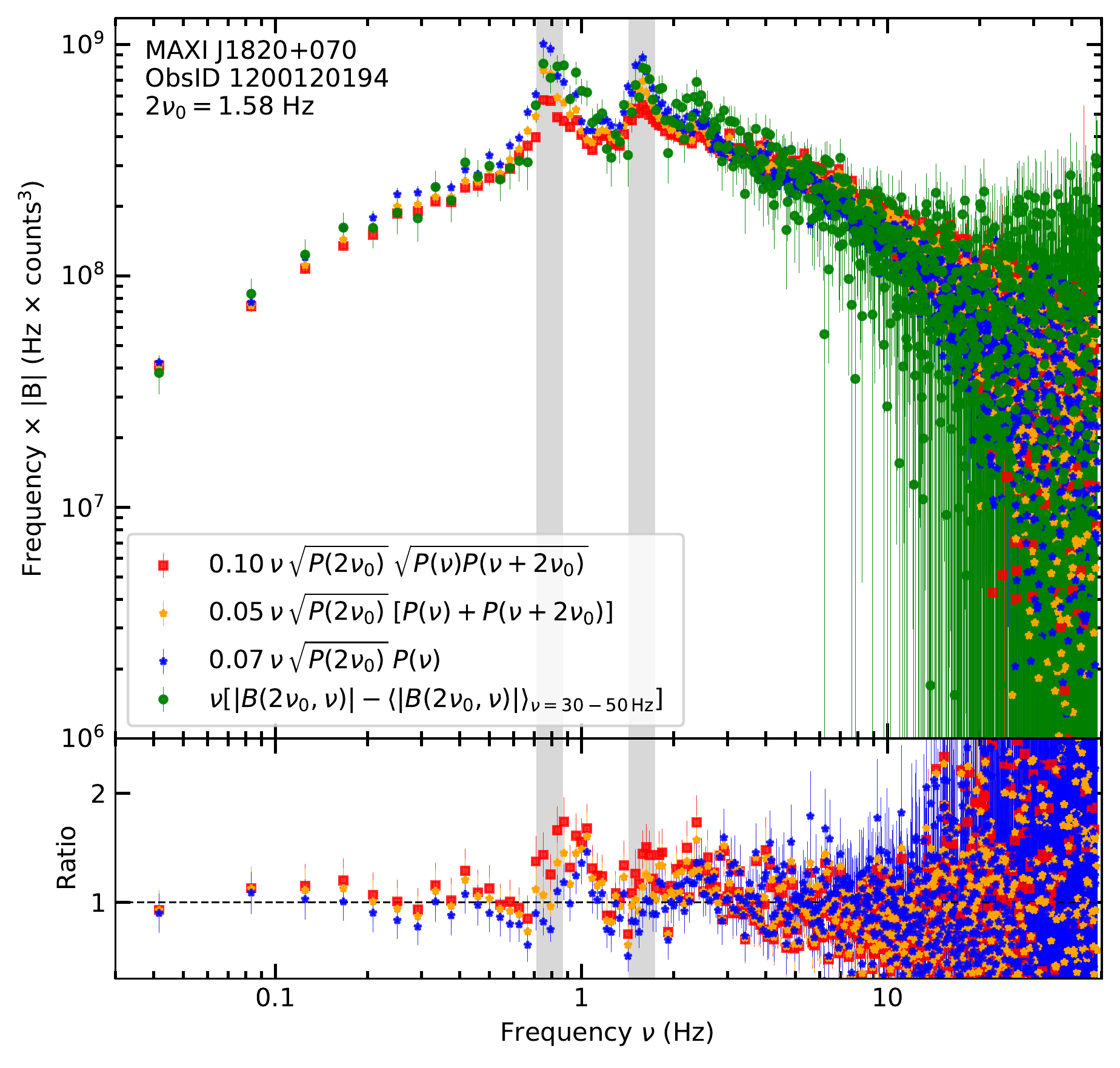}
    \vspace{-2mm}
    \caption{Bispectral modulus along the second harmonic ridge of MAXI~J1820$+$070. The plotted quantities and color scheme are the same as those used in Fig.~\ref{fig:harmonic_ridge_J1727}.}
    \label{fig:second_ridge_J1820}
\end{figure}

Fig.~\ref{fig:second_ridge_J1820} shows the bispectral modulus along the second harmonic ridge of MAXI~J1820$+$070. All three comparison terms broadly reproduce the overall frequency dependence of the measured bispectral modulus. Around the fundamental QPO, the measured bispectral modulus exceeds the comparison terms, with the most positive residual seen for the Fourier-power triplet.

\subsection{Hypotenuse coupling at \(\nu_1+\nu_2=\nu_0\)}
\label{subsec:fundamental-hypotenuse}

\begin{figure*}
    \centering
    \includegraphics[width=0.33\textwidth]{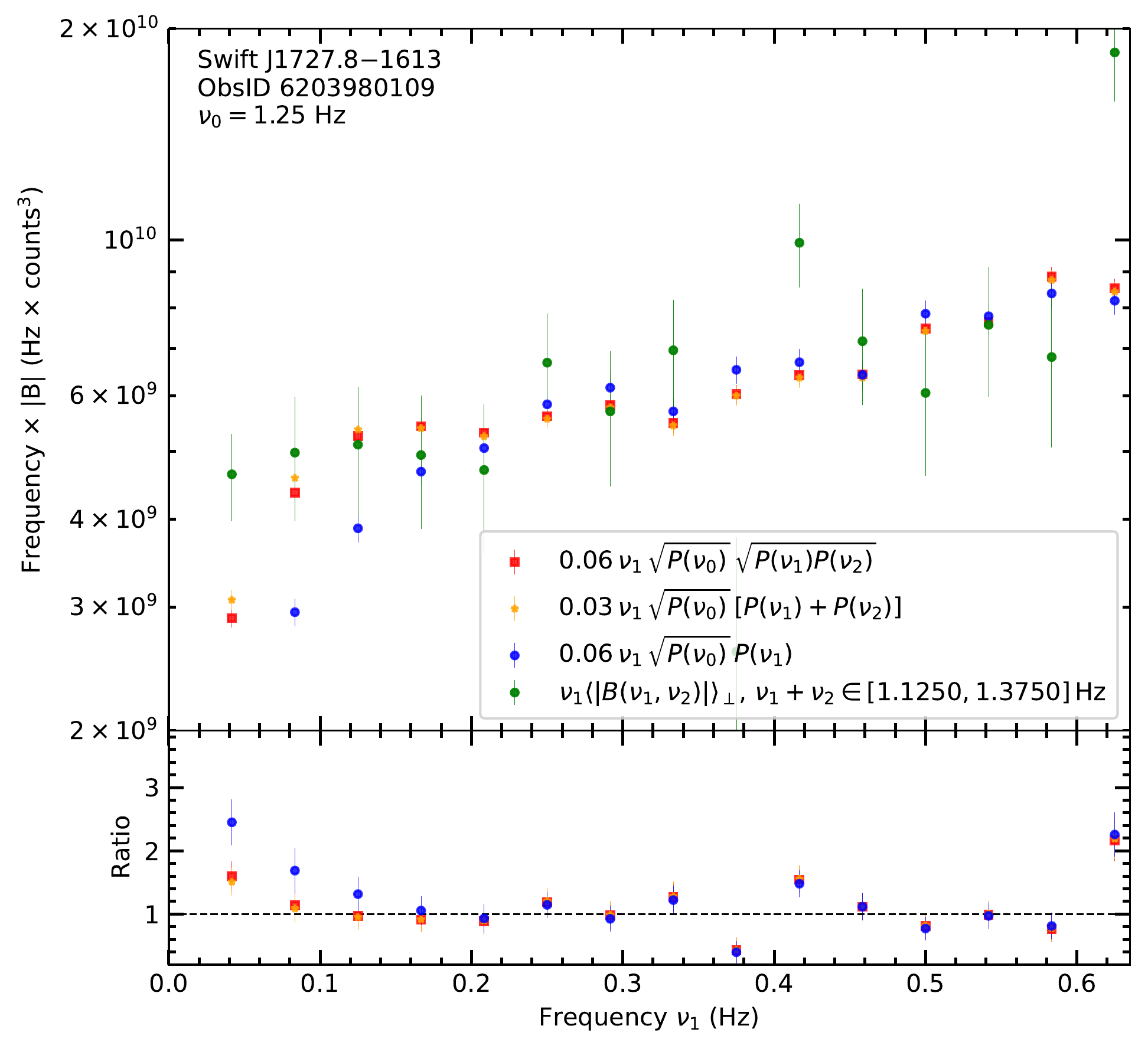}
    \includegraphics[width=0.33\textwidth]{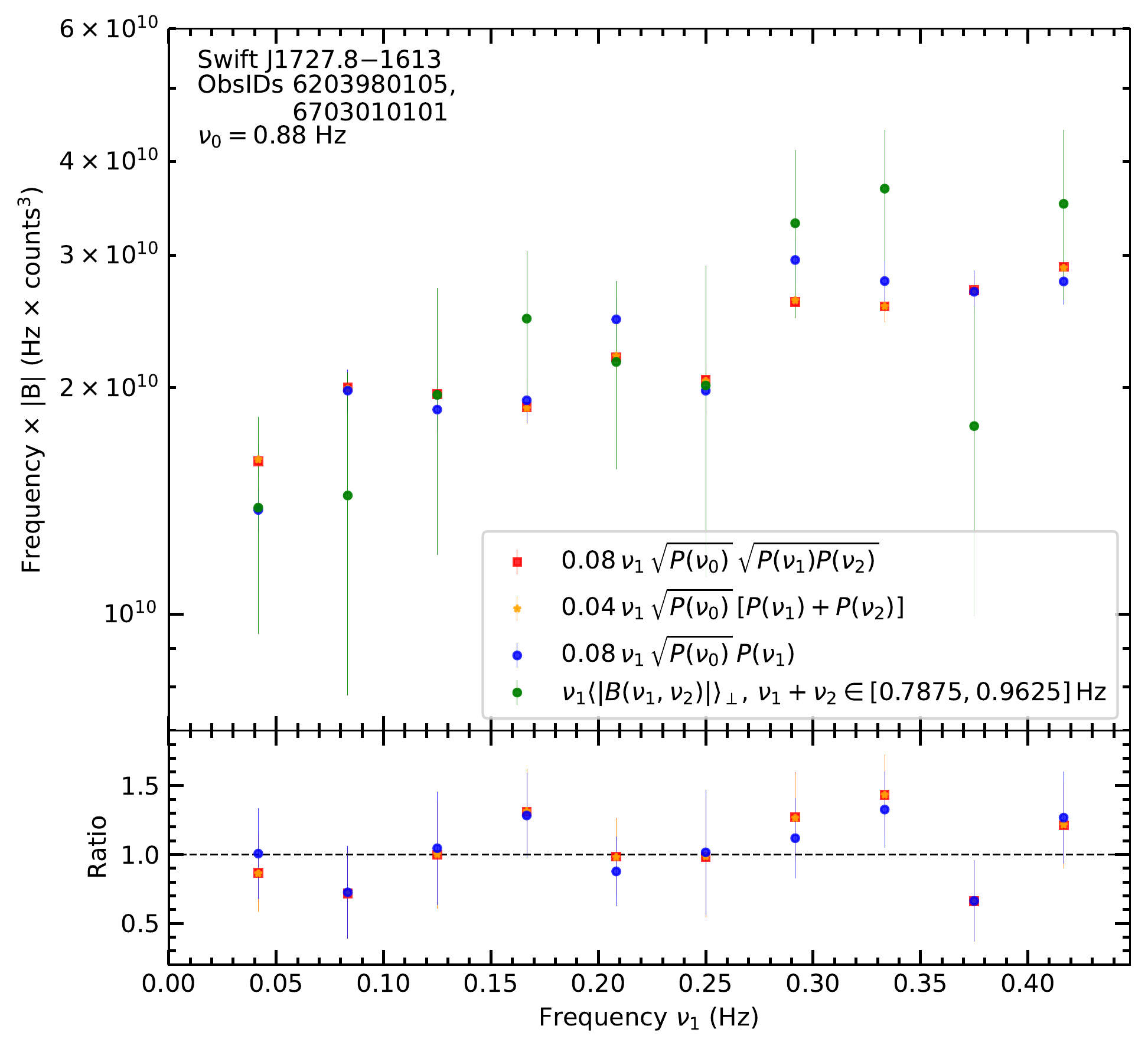}
    \includegraphics[width=0.33\textwidth]{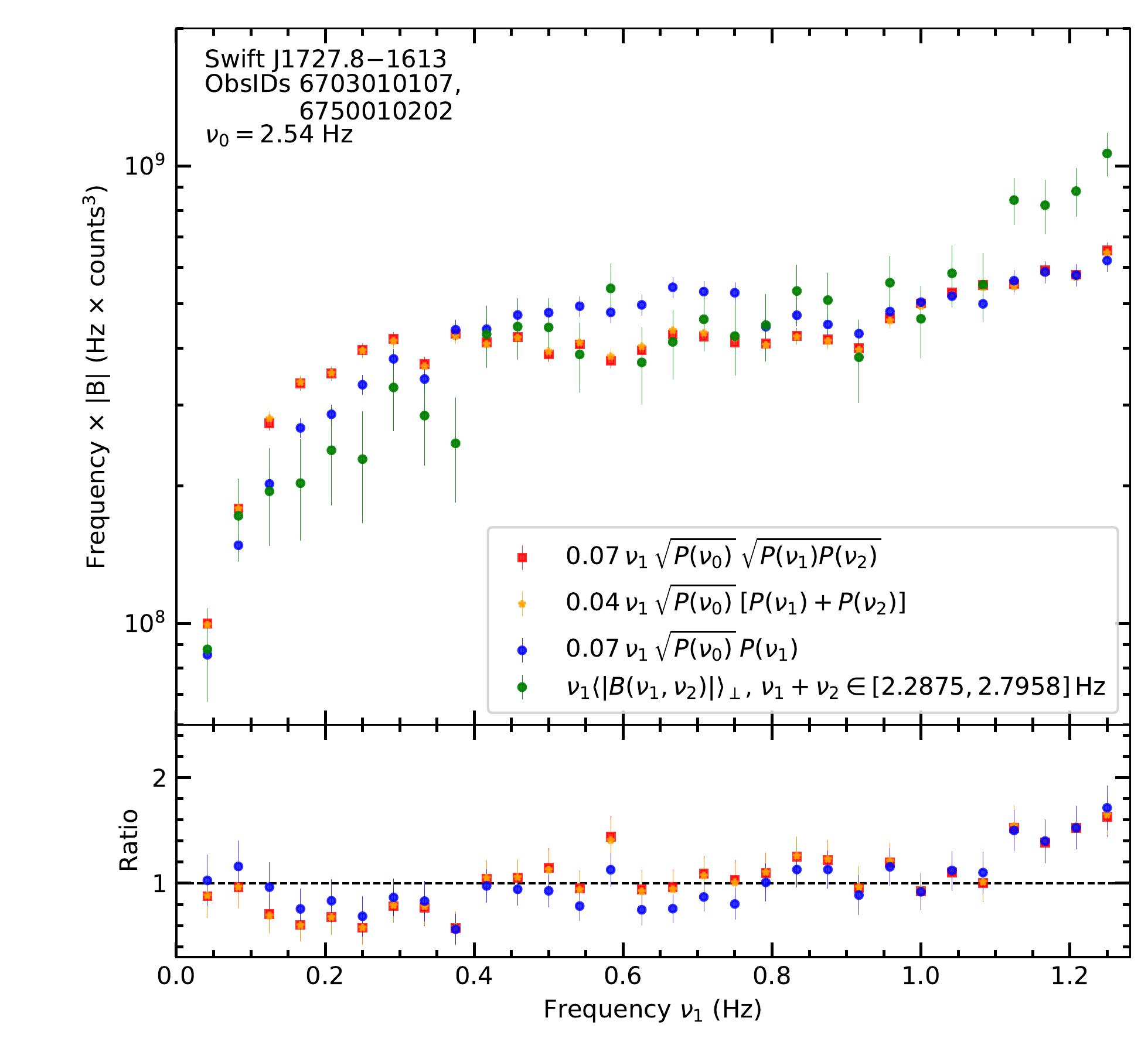}
    \vspace{-2mm}
    \caption{Modulus of the bispectra along the hypotenuse of the fundamental QPO for the three NICER observations of Swift~J1727.8$-$1613. The green points show the measured bispectral modulus, \(|B(\nu_1,\nu_2)|\), with \(\nu_1+\nu_2=\nu_0\). Only the part of the hypotenuse with \(\nu_1<\nu_2\) is shown. The red points show the scaled Fourier-power triplet, \(\sqrt{P(\nu_0)}\sqrt{P(\nu_1)P(\nu_2)}\). The orange points show the scaled modulation-model prediction, \(\sqrt{P(\nu_0)}[P(\nu_1)+P(\nu_2)]\) (see Eq.~\ref{eq:diagonal_mag}). The blue points show the scaled power-spectrum term, \(\sqrt{P(\nu_0)}P(\nu_1)\).
    In each plot the lower panels show the ratios between the bispectral modulus and the three quantities in the upper panels.}
    \label{fig:qpo_hypo_J1727}
\end{figure*}

\begin{figure}
    \centering
    \includegraphics[width=0.4\textwidth]{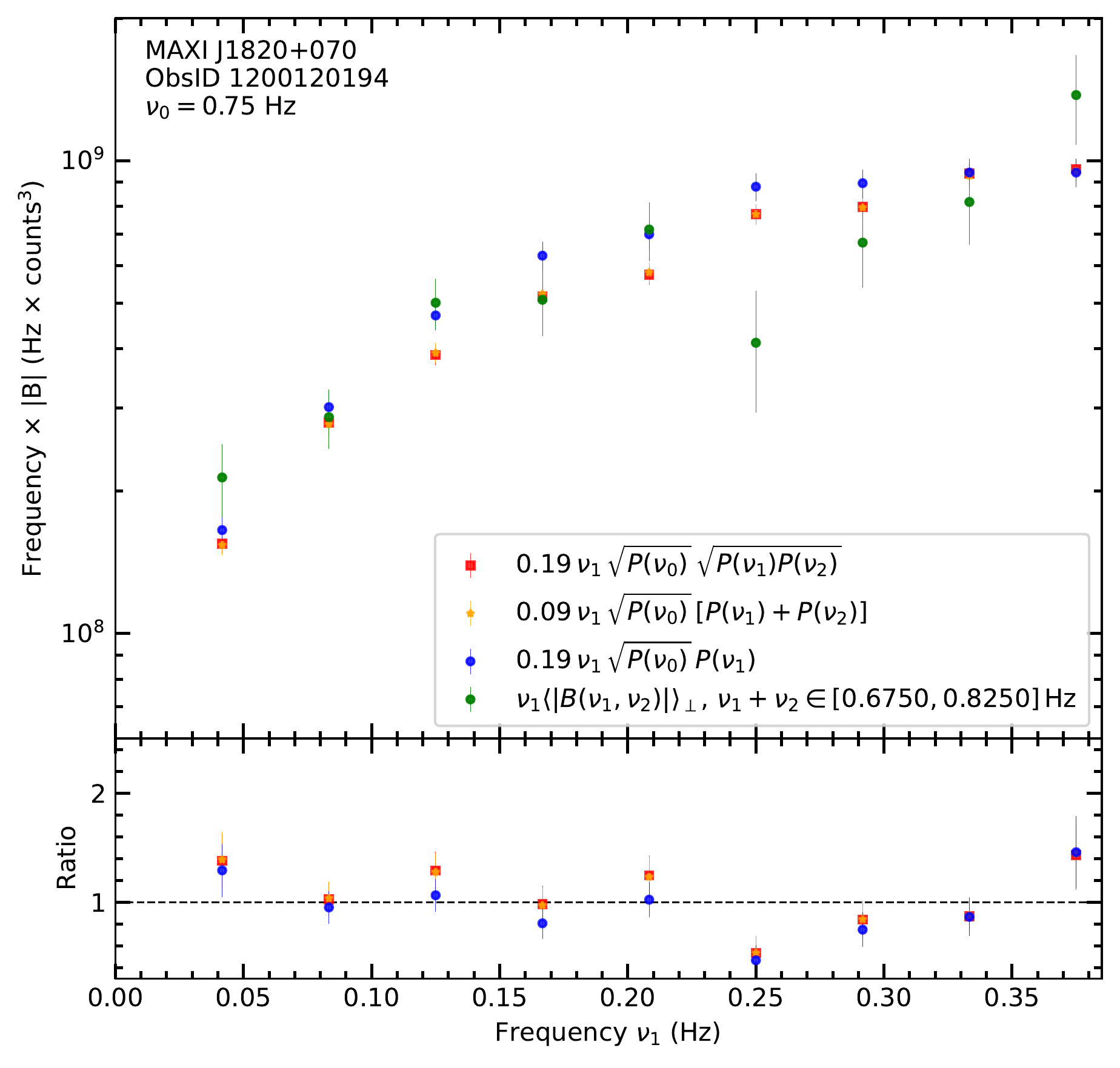}
    \vspace{-2mm}
    \caption{Bispectral modulus along the fundamental QPO hypotenuse of MAXI~J1820$+$070. The plotted quantities and color scheme are the same as those used in Figs.~\ref{fig:qpo_hypo_J1727}.}
    \label{fig:qpo_hypo_J1820}
\end{figure}

Fig.~\ref{fig:qpo_hypo_J1727} shows the bispectral modulus along the hypotenuse associated with the fundamental QPO for the same three NICER observations of Swift~J1727.8$-$1613. 
The bispectral modulus is extracted from the two-dimensional bispectral-modulus map by averaging perpendicular to the fundamental QPO hypotenuse over a frequency width set by \(W_1=0.2\), without binning along the hypotenuse line.
We show only the part of the hypotenuse satisfying \(\nu_1\leq\nu_2\). 
Over most of the frequency range, the measured bispectral modulus is broadly reproduced by all three scaled comparison terms, which have nearly identical frequency dependence.
Fig.~\ref{fig:qpo_hypo_J1820} shows the bispectral modulus and biphase along the fundamental QPO hypotenuse of MAXI~J1820$+$070.
As in Swift~J1727.8$-$1613, the bispectral-modulus comparison alone does not distinguish among these terms.

\subsection{Hypotenuse coupling at \(\nu_1+\nu_2=2\nu_0\)}
\label{subsec:harmonic-hypotenuse}

\begin{figure*}
    \centering
    \includegraphics[width=0.33\textwidth]{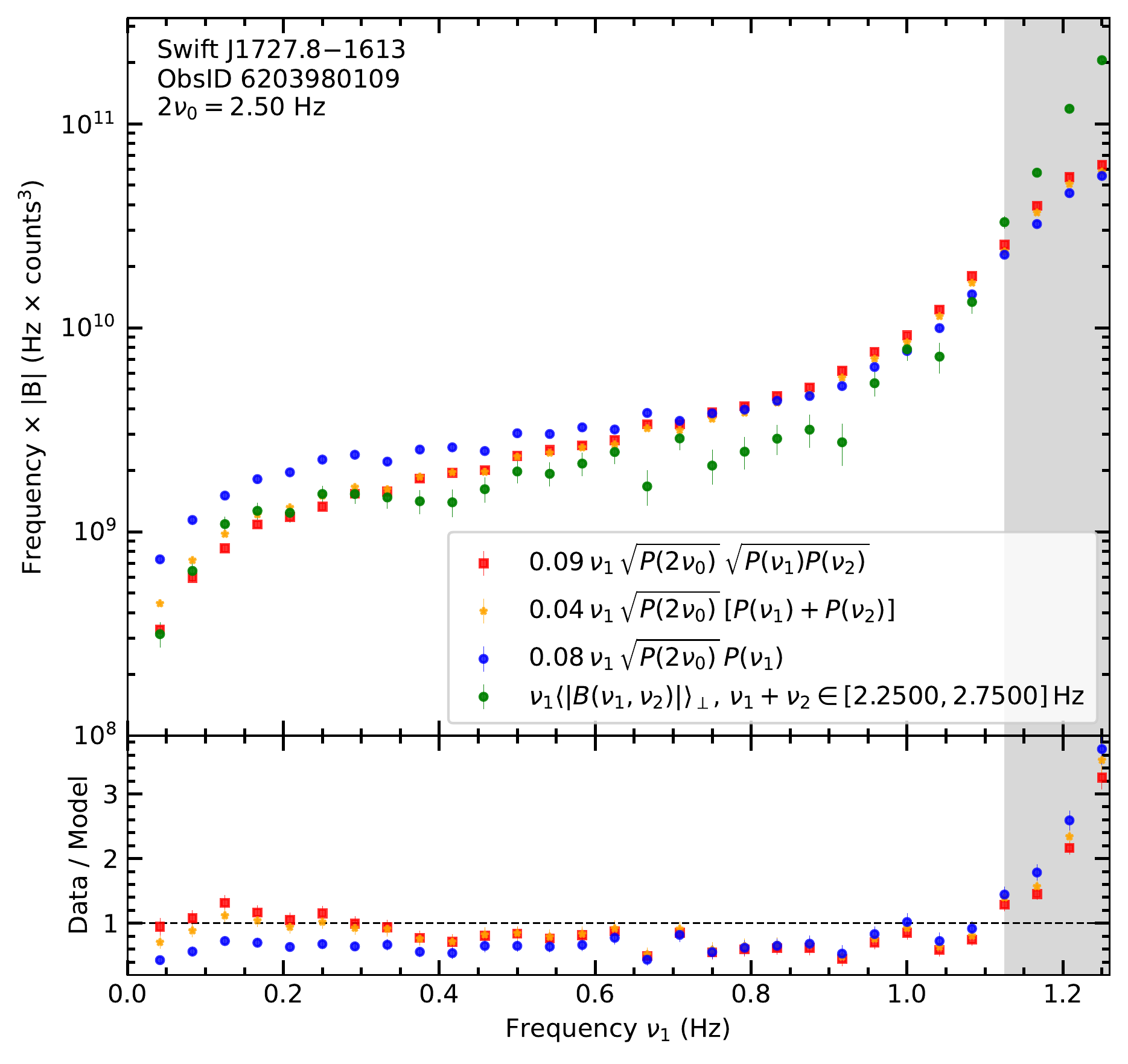}
    \includegraphics[width=0.33\textwidth]{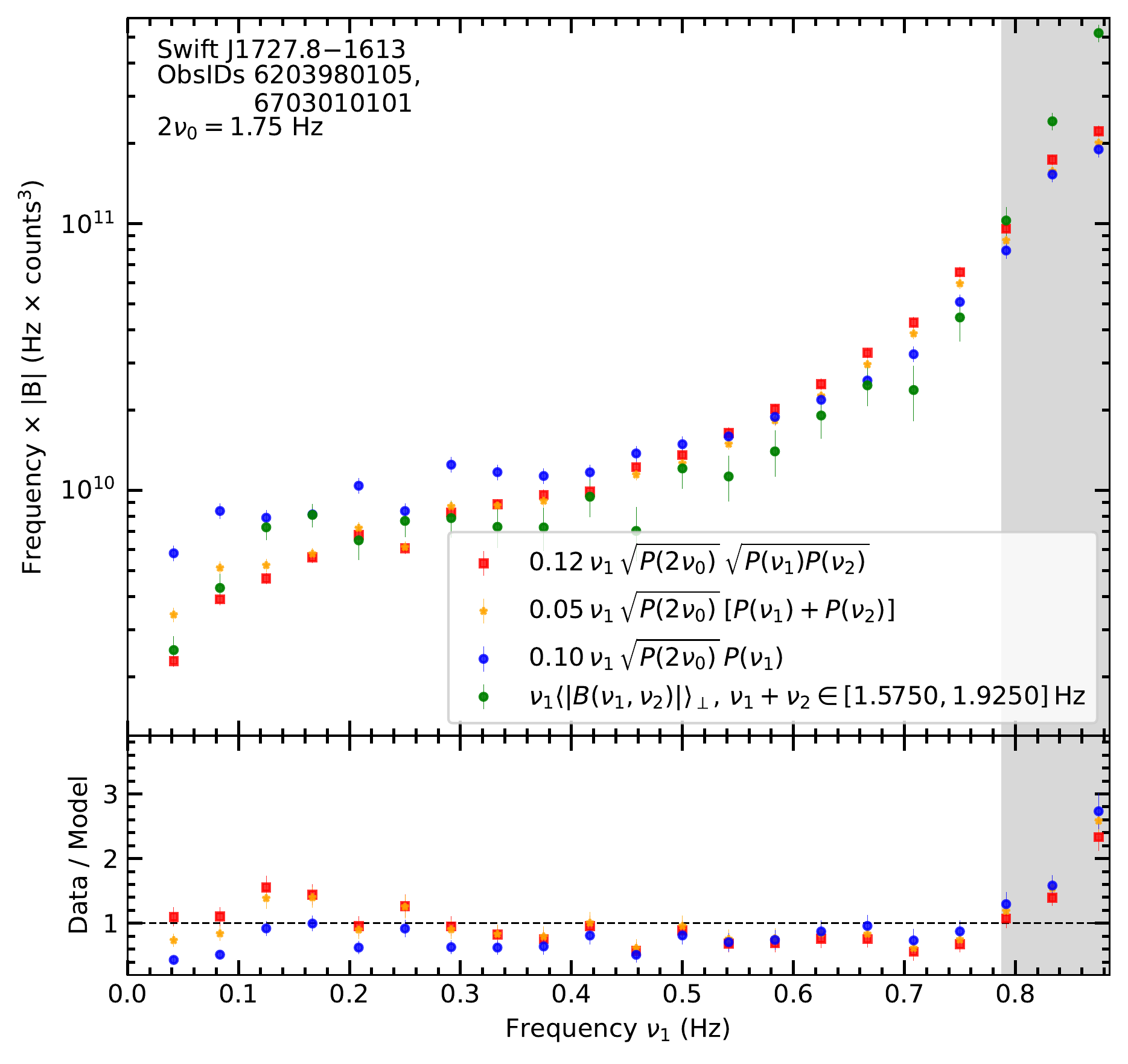}
    \includegraphics[width=0.33\textwidth]{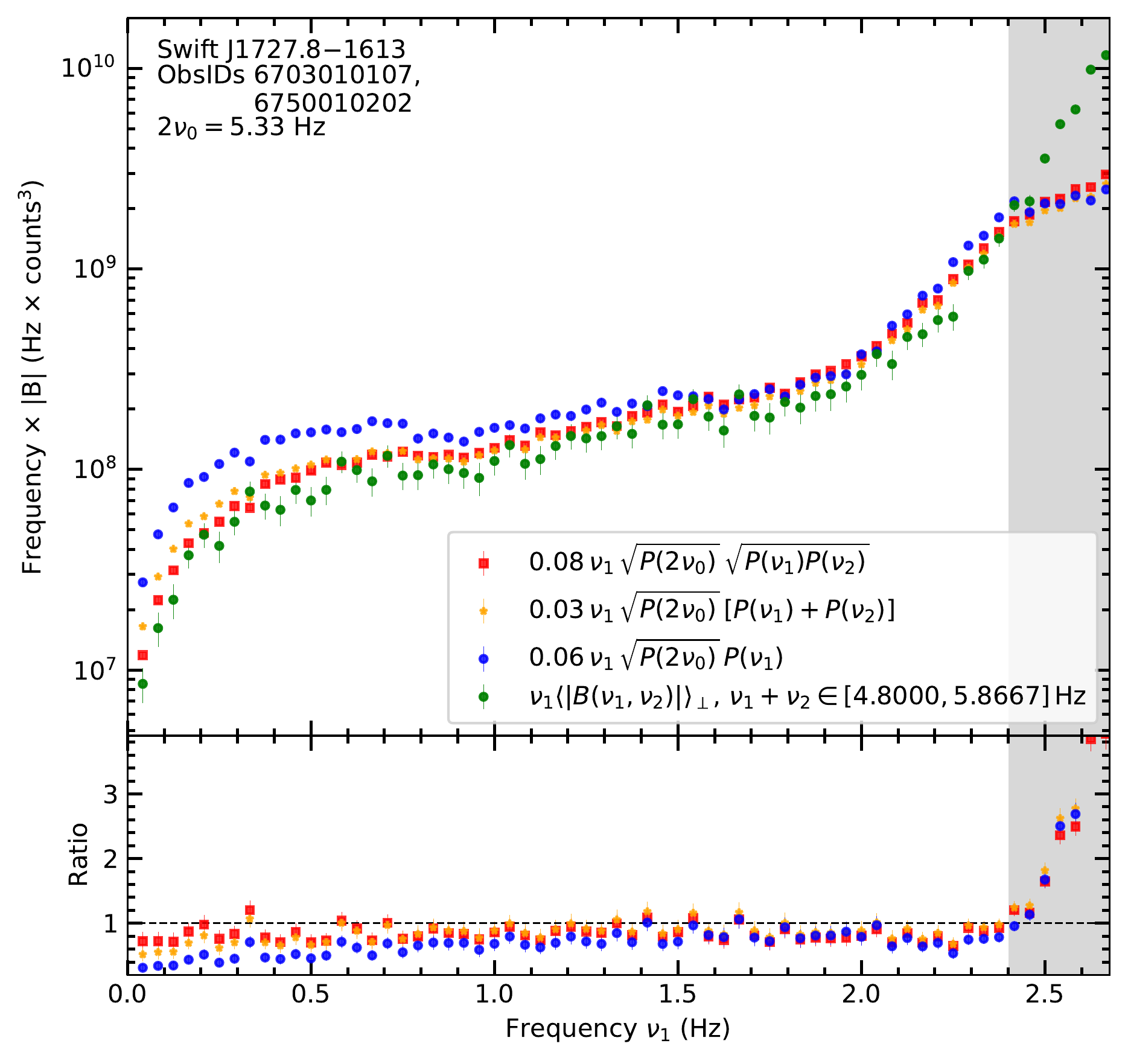}
    \vspace{-2mm}
    \caption{
    Modulus of the bispectra along the hypotenuse of the second harmonic for the three NICER observations of Swift~J1727.8$-$1613. The green points show the measured bispectral modulus, \(|B(\nu_1,\nu_2)|\), with \(\nu_1+\nu_2=2\nu_0\). Only the part of the hypotenuse with \(\nu_1\leq\nu_2\) is shown. The red points show the scaled Fourier-power triplet, \(\sqrt{P(2\nu_0)}\sqrt{P(\nu_1)P(\nu_2)}\). The orange points show the scaled modulation-model prediction, \(\sqrt{P(2\nu_0)}[P(\nu_1)+P(\nu_2)]\) (see Eq.~\ref{eq:harmonic_hypo_mag}). The blue points show the scaled power-spectrum term, \(\sqrt{P(2\nu_0)}P(\nu_1)\).
    In each plot the lower panels show the ratios between the bispectral modulus and the three quantities in the upper panels.
    The gray shaded region highlighs the frequency ranges of the fundamental QPO.}
    \label{fig:harmonic_hypo_J1727}
\end{figure*}

\begin{figure}
    \centering
    \includegraphics[width=0.4\textwidth]{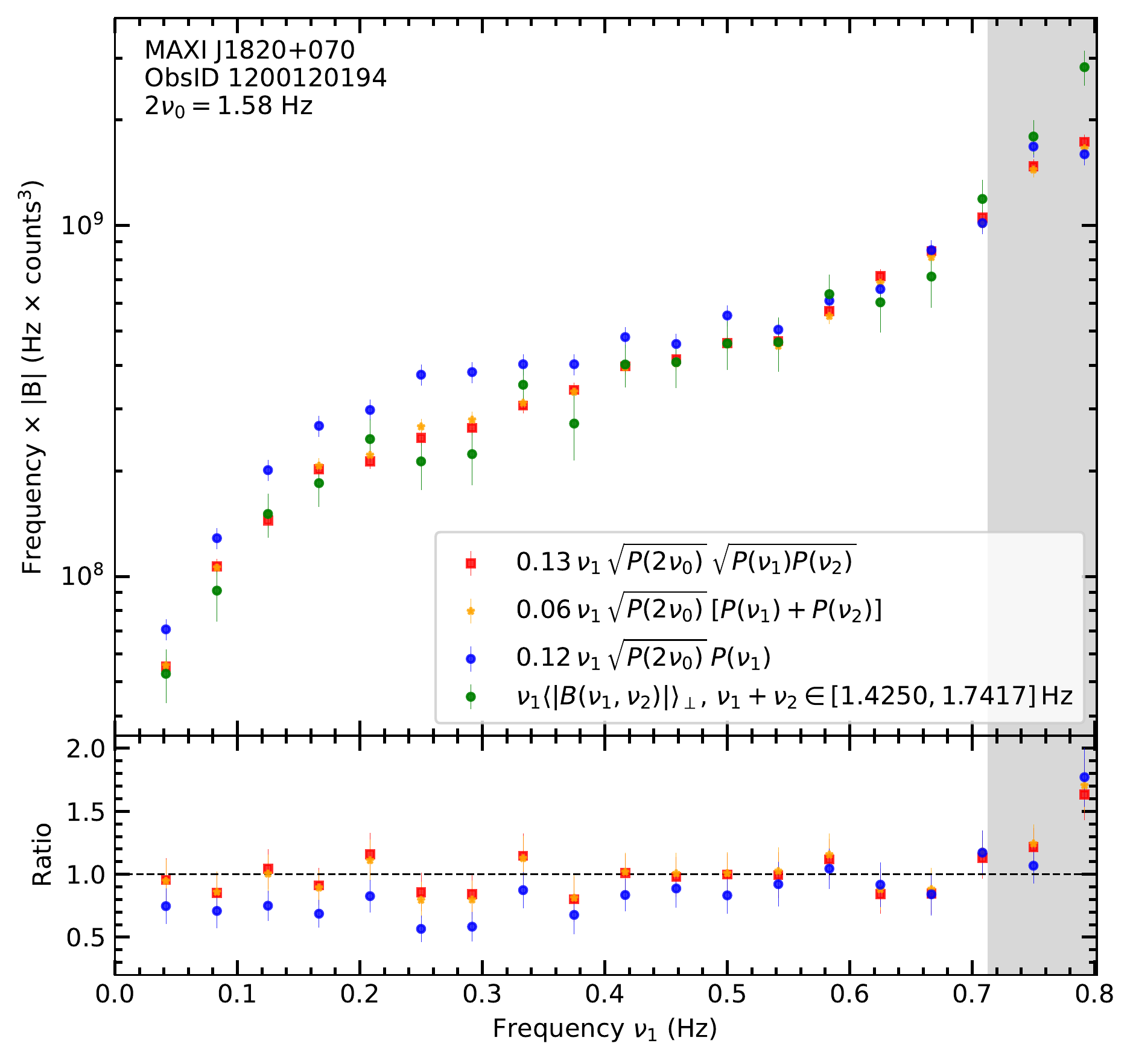}
    \vspace{-2mm}
    \caption{Bispectral modulus along the second harmonic hypotenuse of MAXI~J1820$+$070. The plotted quantities and color scheme are the same as those used in Figs.~\ref{fig:harmonic_hypo_J1727}.}
    \label{fig:second_hypo_J1820}
\end{figure}

Fig.~\ref{fig:harmonic_hypo_J1727} shows the bispectral modulus along the hypotenuse associated with the second harmonic for the same three NICER observations of Swift~J1727.8$-$1613. Only the part of the hypotenuse satisfying \(\nu_1\leq\nu_2\) is shown. In all three observations, the measured bispectral modulus increases with increasing \(\nu_1\).
A pronounced excess is observed near the upper end of the sampled frequency range, particularly as \(\nu_1\) approaches \(\nu_0\). This region corresponds to frequency pairs close to \((\nu_1,\nu_2)=(\nu_0,\nu_0)\) highlighted by the gray region in Fig.~\ref{fig:harmonic_hypo_J1727}, where the second-harmonic hypotenuse overlaps with the fundamental QPO ridge, as seen in the bispectral-modulus and biphase maps presented in Section~\ref{subsec:bispectral-maps}. The simple comparison with the isolated hypotenuse prediction is therefore no longer applicable in this region.
Restricting the comparison to the low-frequency region and the broad minimum, the scaled Fourier-power triplet, the scaled modulation-model prediction, and the scaled power-spectrum term reproduce the overall trend of the measured bispectral modulus.
Fig.~\ref{fig:second_hypo_J1820} shows the bispectral modulus along the second harmonic hypotenuse of MAXI~J1820$+$070. 
The bispectral modulus increases as \(\nu_1\) increases, similar to the behavior observed in Swift~J1727.8$-$1613. All three comparison terms reproduce the measured bispectral modulus below the frequencies at which the contribution from the fundamental QPO ridge becomes dominant.

\section{DISCUSSION}
\label{sec:DISCUSSION}

We developed a mathematical multiplicative modulation model in which a QPO-phase-locked modulator acting on the underlying variability naturally produces the QPO-related ridges and hypotenuses in the bispectral maps. 
While previous studies identified the ridges and hypotenuses as distinct coupling features, our model shows that each ridge and its corresponding hypotenuse could arise together from the same process and have opposite-sign biphases.
We applied the bispectral analysis to NICER observations of Swift~J1727.8$-$1613 and MAXI~J1820$+$070. 
Although some of these structures are weak or difficult to identify in the bicoherence maps, the biphase maps provide a complementary diagnostic by revealing clearer structural trends, even where the bispectral signal is weak.
Both sources show the ridge and hypotenuse components characteristic of a web-like pattern, broadly consistent with the multiplicative-modulation model, with their biphases qualitatively showing the opposite signs.
This raises the possibility that some previously reported ``cross'' or ``hypotenuse'' patterns~\citep{2022MNRAS.514.1720A} may contain complementary components that are too weak to be clearly detected in the bicoherence alone, motivating a re-analysis of these classifications using the complex bispectrum.

We further compared the measured bispectral modulus with the frequency dependence predicted by the modulation model. 
The power spectrum closely reproduces the observed trends in the bispectral modulus, while localized residuals remain around the fundamental QPO and harmonic frequencies. These residuals occur where different bispectral structures overlap, such that the measured bispectral modulus cannot be described by an isolated ridge or hypotenuse prediction.

\subsection{Multiplicative coupling could be a possible origin of the QPO-related bispectral structures}

In our model, the observed variability is written as $x(t)=[1+m(t)]s(t)$, where a modulator phase-locked to the QPO acts on the intrinsic variability $s(t)$. 
The multiplication in the time domain corresponds to frequency mixing (convolution) in the Fourier domain: each Fourier component of $s(t)$ is copied to sidebands separated by the modulation frequency. 
The copied component retains the phase of the original variability while acquiring the phase of the modulator. 
Consequently, Fourier components that are otherwise mutually incoherent can acquire a stable third-order phase relation and hence produce a non-zero bispectrum. 
In this sense, the QPO-related bispectral structures do not require the broadband Fourier components themselves to be intrinsically phase coupled.

The main result of this work is that the QPO-related bispectral structures, ridges and hypotenuses, could arise naturally from the same multiplicative coupling between a modulator that is phase-locked to the QPO and the underlying variability. 
These structures have often been treated as different types of coupling, whereas our model shows that they can be two manifestations of the same process. 
If the ridge and hypotenuse arise from the same mechanism, their relative strengths and phases should not be completely independent. 
In particular, our model predicts that the ridge and its corresponding hypotenuse should have opposite-sign biphases.

The main features in the observed bispectral maps are consistent with the predictions of this model. 
In Swift~J1727.8$-$1613, the bispectral signal is concentrated primarily along the ridges and hypotenuses associated with the fundamental QPO and its second harmonic, producing the web-like pattern. 
Similar QPO-related structures are also observed in MAXI~J1820$+$070. 
The observed biphases show the qualitative opposite-sign tendency between the ridge and corresponding hypotenuse, consistent with the prediction of the model.
A detailed investigation of the biphase structures of these components will be presented in a forthcoming paper.

Our model cannot determine whether the QPO modulates the broadband variability or vice versa. 
This is because multiplication is commutative, and an equivalent mathematical description can always be constructed in which the broadband variability modulates the QPO.
However, we propose a possible physical picture in which the physical or geometrical properties of the corona that produces the QPO could vary quasi-periodically.
Such variations have been considered in physical models of QPOs, including changes in the seed-photon temperature and coronal electron temperature in the \texttt{vkompth} model~\citep{2020MNRAS.492.1399K,2022MNRAS.515.2099B}, and geometrical changes associated with Lense-Thirring precession~\citep{2009MNRAS.397L.101I}.
These variations could also modify the response of the corona to the underlying broadband variability as it propagates through the corona, thereby changing its observed amplitude. 
In this picture, the non-radiative modulator \(m(t)\) describes the variations of the coronal response and is naturally phase-locked to the radiative QPO, \(q(t)\), produced by the same corona.
The coronal response could not be instantaneous or identical at all photon energies, and additional propagation effects could further reshape the bispectral structures. Consequently, the strengths and biphases of the ridges and hypotenuses may depend on photon energy.

\subsection{What shapes the bispectral modulus?}
\label{subsec:bispectral_magnitude}

The multiplicative-modulation model predicts that the frequency dependence of the bispectral modulus is largely determined by the independently measured power spectrum. After an overall scaling, the power spectrum broadly reproduces the frequency dependence of the measured bispectral modulus along both the ridges and the hypotenuses. 
The fact that the different comparison functions reproduce the observed frequency dependence does not necessarily mean that they describe the same physical coupling mechanism. In particular, when \(P(\nu+\nu_0)\) is much smaller than \(P(\nu)\), the prediction of the modulation model becomes nearly indistinguishable from the one-sided approximation. Therefore, the agreement between the bispectral modulus and the power spectrum supports a connection between second- and third-order variability statistics, but does not by itself uniquely identify the coupling mechanism. A more specific feature of our model is that the same modulation produces both the ridge and the corresponding hypotenuse, with a predicted relation between their biphases.

The largest deviations occur near the fundamental QPO and its second harmonic, where different bispectral structures intersect. At these frequencies, the measured bispectrum contains contributions from more than one structure, which must be added as complex quantities before calculating the modulus. As discussed in Section~\ref{subsec:intrinsic_qpo_harmonic} and Appendix~\ref{app:intersections}, the resulting modulus depends on the amplitudes and relative phases of these contributions. This may explain some of the observed deviations, although a quantitative comparison would require modelling the different contributions simultaneously.

\subsection{Is the QPO-related bispectral structure always a web?}

Previous studies of GRS~1915$+$105 identified three characteristic bicoherence patterns, commonly referred to as the ``cross'',
``hypotenuse'', and ``web'' patterns~\citep{2022MNRAS.514.1720A}. 
The observed pattern was also found to evolve with QPO frequency and other source properties. 
The web pattern contains both ridge and hypotenuse features, whereas the cross and hypotenuse patterns are dominated by ridges and hypotenuse features, respectively.

Our results suggest that the absence of a feature in the bicoherence map may not imply the absence of the corresponding coupling. 
In Swift~J1727.8$-$1613, some hypotenuse structures are difficult to identify from the bicoherence alone but become much clearer in the biphase map. 
This is even more apparent in MAXI~J1820$+$070: its bicoherence map does not show a prominent web-like morphology, whereas the biphase reveals coherent opposite-sign structures along the fundamental-QPO ridge and hypotenuse.
Therefore, the biphase map could help reveal weak coupling components that are difficult to identify from the bicoherence map alone, as it could exhibit clearer trends even when the statistical significance of the bispectral signal is low.

The web-like structures are consistent with the predictions of our modulation model, in which a single modulation process produces both the ridges and their corresponding hypotenuses. 
This suggests that the distinction between the ``cross'', ``hypotenuse'', and ``web'' patterns may not reflect different physical coupling mechanisms. 
Instead, changes in the relative strengths of the ridge and hypotenuse components could produce apparent transitions between these patterns without requiring a change in the underlying coupling mechanism.

A systematic reanalysis of sources such as GRS~1915$+$105, using the complex bispectrum in addition to the bicoherence, is therefore needed to test whether the previously identified ``cross'' and ``hypotenuse'' patterns also contain weaker hypotenuse and ridge features, respectively, that are difficult to detect with the bicoherence alone.
Such an analysis would provide a direct test of the prediction of our modulation model that the ridge and hypotenuse are two manifestations of the same multiplicative coupling process.

Our model relies on several simplifying assumptions. First, we assume weak modulation and retain only first-order terms in the modulation amplitudes. 
This allows us to approximate the intrinsic power spectrum by the observed power spectrum. 
Second, we assume that the variability components represented by the different Lorentzians are independent Gaussian stochastic processes, 
so that their intrinsic bispectra and mixed third-order terms vanish under ensemble averaging. 
This assumption may not hold if the broadband variability itself exhibits nonlinear phase coupling, which could introduce additional contributions to the observed bispectrum. 
This assumption applies only to the broadband noise components; intrinsic phase coupling between the fundamental QPO and its harmonics is allowed in our model.
Finally, propagation effects, such as reflection and additional Comptonization, and delayed responses, which are not explicitly included in the model, could also modify the observed bispectral modulus and biphases in an energy-dependent picture. These effects remain to be investigated in future work.

\section{CONCLUSIONS}
\label{sec:CONCLUSIONS}

The main conclusions of this work are:

\begin{enumerate}

\item We developed a mathematical model in which multiplicative coupling between the QPO-phase-locked modulator and the underlying variability naturally produces the QPO-related ridges and hypotenuses in the bispectrum. 
In this model, each ridge and its corresponding hypotenuse are not independent features, but could arise together from the same modulation. 
The model further predicts opposite-sign biphases between each ridge and its corresponding hypotenuse.

\item The bispectral maps of Swift~J1727.8$-$1613 and MAXI~J1820$+$070 show the QPO-related ridge-hypotenuse structures with qualitatively opposite-sign biphases, consistent with the predictions of the model. 
The biphase map provides a complementary diagnostic for identifying weak coupling structures that are difficult to detect from the bicoherence map alone.

\item The power spectrum closely reproduces the frequency dependence of the bispectral modulus along both the ridges and hypotenuses.
Along these particular lines, significant model residuals become apparent around the fundamental QPO and second harmonic frequencies. 
This is expected, as multiple bispectral structures overlap and contribute simultaneously. 

\item These results support the idea that multiplicative coupling could be a possible explanation for the web-like bispectrum. 

In future work, we will investigate the biphase properties in more detail.

\end{enumerate}

{\em Note added at submission:}
During the preparation of this work, we became aware of Larner (2026, submitted concurrently), who independently investigate a similar multiplicative coupling mechanism. Like us, they show that modulation of broadband variability by a QPO and its harmonic can produce the ridge (cross) and hypotenuse structures characteristic of the web-like bispectral pattern, with related biphase signatures. While our work focuses on the analytical predictions for the complex bispectrum, including the relative phase between the observed QPO and the modulator, and their comparison with NICER observations of Swift J1727.8$-$1613 and MAXI J1820$+$070, Larner develops a model that additionally includes quadratic coupling of broadband variability and incoherent dilution. He use these ingredients to investigate transitions between different bicoherence patterns and their possible dependence on inclination.

\section*{Acknowledgements}

We thank the Bispectrum Support Group for helpful discussions. PJ acknowledges support from the China Scholarship Council (CSC 202304910058). FG is a CONICET researcher. FG acknowledges support from CAS-ANSO Fellowship program. FG was partially supported by grant PID2022-136828NB-C42 funded by the Spanish MCIN/AEI/
10.13039/501100011033 and “ERDF A way of making Europe”.

\section*{Data Availability}

The data analysed in this paper are publicly available through the HEASARC Browse service at
\url{https://heasarc.gsfc.nasa.gov/cgi-bin/W3Browse/w3browse.pl}.



\bibliographystyle{mnras}
\bibliography{bispectrum} 




\appendix

\section{Bispectrum at intersections of modulation-induced structures}
\label{app:intersections}

The analytic expressions derived in Sections~\ref{sec:MATHEMATICAL FORMALISM} describe the bispectral contribution associated with an individual ridge or hypotenuse. In the two-dimensional bispectrum, however, these structures intersect at several frequency pairs. At such locations, more than one modulation-induced contribution can survive simultaneously, and the measured bispectrum is their complex sum. 

Including modulation at both the fundamental QPO frequency and its second harmonic, the Fourier transform of the observed signal can be
written as
\begin{align}
X(\nu) = S(\nu)
&+ M_+^{1} S(\nu-\nu_0)
 + M_-^{1} S(\nu+\nu_0)
\nonumber\\
&+ M_+^{2} S(\nu-2\nu_0)
 + M_-^{2} S(\nu+2\nu_0),
\label{eq:app_X_general}
\end{align}
where
\begin{equation}
M_\pm^{1}
\equiv
\frac{u_1}{2}e^{\pm i\psi_{m,1}},
\qquad
M_\pm^{2}
\equiv
\frac{u_2}{2}e^{\pm i\psi_{m,2}}.
\label{eq:app_Mpm}
\end{equation}
The modulation phases satisfy
\begin{equation}
\psi_{m,1}=\psi_1+\Delta_1,
\qquad
\psi_{m,2}=\psi_2+\Delta_2.
\label{eq:app_phase_offsets}
\end{equation}

These definitions are identical to those adopted in Section~\ref{sec:MATHEMATICAL FORMALISM}. The corresponding isolated ridge predictions are proportional to $e^{-i\Delta_1}$ and $e^{-i\Delta_2}$, respectively, whereas the hypotenuse contributions have opposite phase signs.

In the following, we retain terms to first order in $u_1$ and $u_2$. Terms whose phases do not cancel under ensemble averaging are
neglected, consistent with the assumptions used in deriving the isolated ridge and hypotenuse expressions in the main text.

\subsection{Intersection at $(\nu_0,\nu_0)$}
\label{app:intersection_11}

At the fundamental-QPO diagonal,
\begin{equation}
B(\nu_0,\nu_0)
=
\left\langle
X(\nu_0)X(\nu_0)X^*(2\nu_0)
\right\rangle .
\label{eq:app_B11}
\end{equation}
The relevant Fourier amplitudes are
\begin{align}
X(\nu_0) ={}&
S(\nu_0)
+M_-^1S(2\nu_0)
+M_+^1S(0)
\nonumber\\
&+M_-^2S(3\nu_0)
+M_+^2S(-\nu_0),
\label{eq:app_X1}
\end{align}
and
\begin{align}
X^*(2\nu_0) ={}&
S^*(2\nu_0)
+M_+^1S^*(3\nu_0)
+M_-^1S^*(\nu_0)
\nonumber\\
&+M_+^2S^*(4\nu_0)
+M_-^2S^*(0).
\label{eq:app_X2star}
\end{align}

To first order in the modulation amplitudes, the phase-surviving terms are
\begin{align}
B(\nu_0,\nu_0) \simeq{}&
2M_-^1
\left\langle
S(\nu_0)|S(2\nu_0)|^2
\right\rangle
\nonumber\\
&+
M_-^1
\left\langle
S(\nu_0)|S(\nu_0)|^2
\right\rangle
\nonumber\\
&+
2M_+^2
\left\langle
|S(\nu_0)|^2S^*(2\nu_0)
\right\rangle .
\label{eq:app_B11_survive}
\end{align}

Using
\begin{equation}
M_-^1S(\nu_0)
=
\frac{u_1a_1}{4}e^{-i\Delta_1},
\qquad
M_+^2S^*(2\nu_0)
=
\frac{u_2a_2}{4}e^{+i\Delta_2},
\label{eq:app_phase_products}
\end{equation}
we obtain
\begin{align}
B(\nu_0,\nu_0) \simeq{}&
\underbrace{
\frac{u_1a_1}{4}e^{-i\Delta_1}
\left[
P(\nu_0)+2P(2\nu_0)
\right]
}_{B_{\mathrm{ridge},1}}
\nonumber\\
&+
\underbrace{
\frac{u_2a_2}{4}e^{+i\Delta_2}
2P(\nu_0)
}_{B_{\mathrm{hypot},2}} .
\label{eq:app_B11_result}
\end{align}

The first contribution is associated with the fundamental-QPO ridge, while the second is associated with the second-harmonic hypotenuse, since $\nu_0+\nu_0=2\nu_0$. The factors of two arise because the first two Fourier factors in Eq.~(\ref{eq:app_B11}) are identical, so that the corresponding shifted component can be selected from either factor.

If the intrinsic phase relation
\begin{equation}
\psi_2=2\psi_1+\Delta_{12}
\end{equation}
is also present, the zeroth-order term contributes
\begin{equation}
B_{\rm int}(\nu_0,\nu_0)
\simeq
\frac{a_1^2a_2}{8}e^{-i\Delta_{12}},
\label{eq:app_Bint}
\end{equation}
as derived in Section~\ref{subsec:intrinsic_qpo_harmonic}. Thus, the complete contribution at this intersection can be written schematically as
\begin{equation}
B(\nu_0,\nu_0)
=
B_{\mathrm{ridge},1}
+
B_{\mathrm{hypot},2}
+
B_{\rm int},
\label{eq:app_B11_total}
\end{equation}
where the last term is present only if intrinsic fundamental--harmonic phase coupling is included.

\subsection{Intersection of the fundamental and second-harmonic ridges}
\label{app:intersection_12}

We next consider
\begin{equation}
B(\nu_0,2\nu_0)
=
\left\langle
X(\nu_0)X(2\nu_0)X^*(3\nu_0)
\right\rangle .
\label{eq:app_B12}
\end{equation}
The additional Fourier amplitudes required are
\begin{align}
X(2\nu_0) ={}&
S(2\nu_0)
+M_-^1S(3\nu_0)
+M_+^1S(\nu_0)
\nonumber\\
&+M_-^2S(4\nu_0)
+M_+^2S(0),
\label{eq:app_X2}
\end{align}
and
\begin{align}
X^*(3\nu_0) ={}&
S^*(3\nu_0)
+M_+^1S^*(4\nu_0)
+M_-^1S^*(2\nu_0)
\nonumber\\
&+M_+^2S^*(5\nu_0)
+M_-^2S^*(\nu_0).
\label{eq:app_X3star}
\end{align}

The phase-surviving first-order terms give
\begin{align}
B(\nu_0,2\nu_0) \simeq{}&
M_-^1S(\nu_0)
\left[
P(2\nu_0)+P(3\nu_0)
\right]
\nonumber\\
&+
M_-^2S(2\nu_0)
\left[
P(\nu_0)+P(3\nu_0)
\right].
\label{eq:app_B12_survive}
\end{align}
Since
\begin{equation}
M_-^2S(2\nu_0)
=
\frac{u_2a_2}{4}e^{-i\Delta_2},
\label{eq:app_M2S2}
\end{equation}
this becomes
\begin{align}
B(\nu_0,2\nu_0) \simeq{}&
\underbrace{
\frac{u_1a_1}{4}e^{-i\Delta_1}
\left[
P(2\nu_0)+P(3\nu_0)
\right]
}_{B_{\mathrm{ridge},1}}
\nonumber\\
&+
\underbrace{
\frac{u_2a_2}{4}e^{-i\Delta_2}
\left[
P(\nu_0)+P(3\nu_0)
\right]
}_{B_{\mathrm{ridge},2}} .
\label{eq:app_B12_result}
\end{align}

Hence,
\begin{equation}
B(\nu_0,2\nu_0)
=
B_{\mathrm{ridge},1}
+
B_{\mathrm{ridge},2}.
\label{eq:app_B12_total}
\end{equation}

Because the bispectrum is symmetric under interchange of its first two frequencies,
\begin{equation}
B(\nu_1,\nu_2)=B(\nu_2,\nu_1),
\end{equation}
the same result applies to
\begin{equation}
B(2\nu_0,\nu_0)
=
B(\nu_0,2\nu_0).
\label{eq:app_B21}
\end{equation}
Thus, $(\nu_0,2\nu_0)$ and $(2\nu_0,\nu_0)$ represent the same physical intersection between the fundamental-QPO and second-harmonic ridges.

\subsection{The second-harmonic diagonal at $(2\nu_0,2\nu_0)$}
\label{app:intersection_22}

Finally, consider
\begin{equation}
B(2\nu_0,2\nu_0)
=
\left\langle
X(2\nu_0)X(2\nu_0)X^*(4\nu_0)
\right\rangle ,
\label{eq:app_B22}
\end{equation}
where
\begin{align}
X^*(4\nu_0) ={}&
S^*(4\nu_0)
+M_+^1S^*(5\nu_0)
+M_-^1S^*(3\nu_0)
\nonumber\\
&+M_+^2S^*(6\nu_0)
+M_-^2S^*(2\nu_0).
\label{eq:app_X4star}
\end{align}

At first order, the surviving second-harmonic modulation terms are
\begin{align}
B(2\nu_0,2\nu_0) \simeq{}&
2M_-^2
\left\langle
S(2\nu_0)|S(4\nu_0)|^2
\right\rangle
\nonumber\\
&+
M_-^2
\left\langle
S(2\nu_0)|S(2\nu_0)|^2
\right\rangle ,
\label{eq:app_B22_survive}
\end{align}
which gives
\begin{equation}
B(2\nu_0,2\nu_0)
\simeq
\frac{u_2a_2}{4}e^{-i\Delta_2}
\left[
P(2\nu_0)+2P(4\nu_0)
\right].
\label{eq:app_B22_result}
\end{equation}

A hypotenuse passing through $(2\nu_0,2\nu_0)$ would instead require modulation at $4\nu_0$, which is not included in the present model.
An intrinsic contribution may nevertheless arise if the second harmonic is intrinsically phase coupled to a component at $4\nu_0$. Defining
\begin{equation}
\psi_4 = 2\psi_2 + \Delta_{24},
\end{equation}
the corresponding intrinsic bispectrum is
\begin{equation}
B_{\rm int,24}(2\nu_0,2\nu_0)
\simeq
\frac{a_2^2a_4}{8}e^{-i\Delta_{24}}.
\end{equation}
Thus, to leading order,
\begin{equation}
B(2\nu_0,2\nu_0)
=
B_{\mathrm{ridge},2}
+
B_{\rm int,24}.
\label{eq:app_B22_total}
\end{equation}

\subsection{Implications for ridge measurements}
\label{app:intersection_implications}

The relevant special frequency pairs can therefore be summarized as
\begin{align}
(\nu_0,\nu_0)
&:\quad R_1\cap H_2
\quad [+\,I_{12}],
\nonumber\\
(\nu_0,2\nu_0)
&:\quad R_1\cap R_2,
\nonumber\\
(2\nu_0,\nu_0)
&:\quad R_2\cap R_1,
\nonumber\\
(2\nu_0,2\nu_0)
&:\quad R_2\ [+\,I_{24}].
\label{eq:app_intersection_summary}
\end{align}
Here $R_1$ and $R_2$ denote the fundamental and second-harmonic ridges, respectively, $H_2$ denotes the second-harmonic hypotenuse.
The terms $I_{12}$ and $I_{24}$ denote the intrinsic couplings between the fundamental and second harmonic, and between the second and fourth harmonics, respectively.

\section{Comparison of Bispectral Averaging Methods}
\label{lab:Averaging}

\begin{figure}
    \centering
    \includegraphics[width=0.4\textwidth]{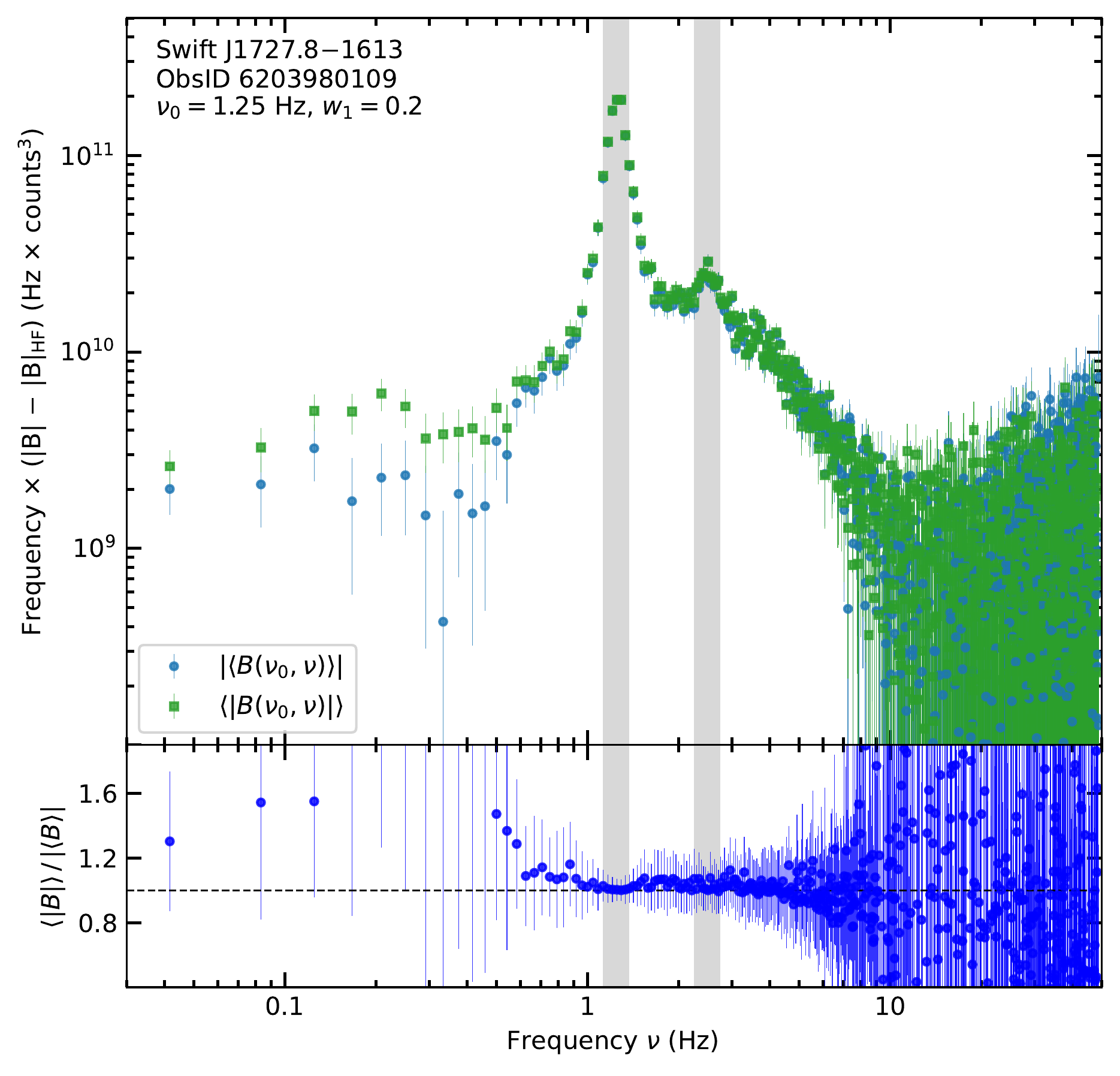}
    \vspace{-2mm}
    \caption{
    Comparison of the bispectral modulus obtained by taking the modulus after averaging the complex bispectrum, $|\langle B(\nu_0,\nu)\rangle|$, and by averaging the modulus directly, $\langle|B(\nu_0,\nu)|\rangle$, along the fundamental-QPO ridge for the NICER observation of Swift~J1727.8$-$1613 with $\nu_0=1.25$~Hz. The lower panel shows the ratio $\langle|B|\rangle/|\langle B\rangle|$.
    }
    \label{fig:average}
\end{figure}

Figure~\ref{fig:average} compares the two methods of averaging the bispectrum across the fundamental-QPO ridge. 
In the first method, the complex bispectrum is averaged before taking its modulus,
$|\langle B(\nu_0,\nu)\rangle|$ (blue), 
whereas in the second method, the modulus is calculated first and then averaged,
$\langle|B(\nu_0,\nu)|\rangle$ (green).
The lower panel shows the ratio of the results obtained using these two methods.
The two estimates are nearly identical over most of the frequency range, including around the QPO-related peaks and the main broadband structure, indicating that the choice of averaging method does not significantly affect the measured bispectral-modulus profile.
At the lowest frequencies, however, $|\langle B\rangle|$ tends to be smaller than $\langle|B|\rangle$.
This difference may be associated with the overlap between the ridge and hypotenuse components in this region.
As shown in Fig.~\ref{fig:bic_bis_bip_J1727_1.25Hz}, the biphase varies more strongly across the overlapping region.
Consequently, averaging the complex bispectrum over the ridge band leads to partial cancellation between contributions with different phases, reducing $|\langle B\rangle|$.
In contrast, $\langle|B|\rangle$ is insensitive to this phase dispersion because the modulus is calculated before averaging.

\section{Bispectral modulus for different averaging widths}
\label{lab:bispectral modulus for different Q}

\begin{figure}
    \centering
    \includegraphics[width=0.4\textwidth]{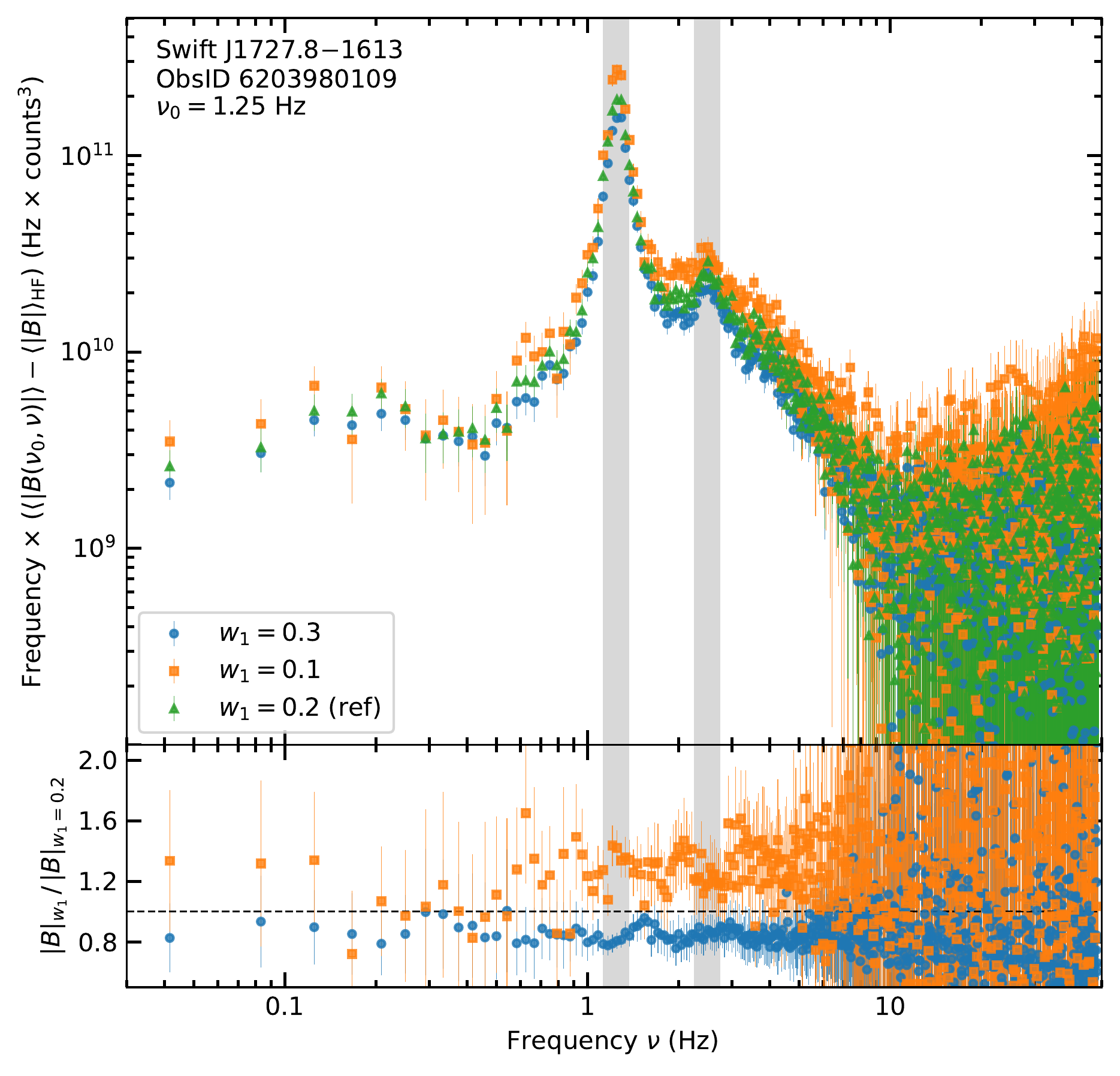}
    \vspace{-2mm}
    \caption{
    Upper panel: bispectral modulus along the fundamental-QPO ridge for the NICER observation of Swift~J1727.8$-$1613 with $\nu_0=1.25$~Hz, calculated using different averaging frequency widths set by $W_1$.
    Lower panel: Ratios relative to the $W_1=0.2$ result.}
    \label{fig:Q}
\end{figure}

Fig.~\ref{fig:Q} shows the bispectral modulus along the fundamental-QPO ridge for the NICER observation of Swift~J1727.8$-$1613 with $\nu_0=1.25$~Hz, calculated using different averaging frequency widths set by $W_1$.
The overall frequency dependence remains similar for different averaging frequency widths, with changes mainly affecting the normalization.
The ratios relative to $W_1=0.2$ further confirm that the bispectral modulus shape depends only weakly on the adopted width.

\section{Dynamical power spectrum}
\label{lab:Dynamical power spectrum}

\begin{figure}
    \centering
    \includegraphics[width=0.4\textwidth]{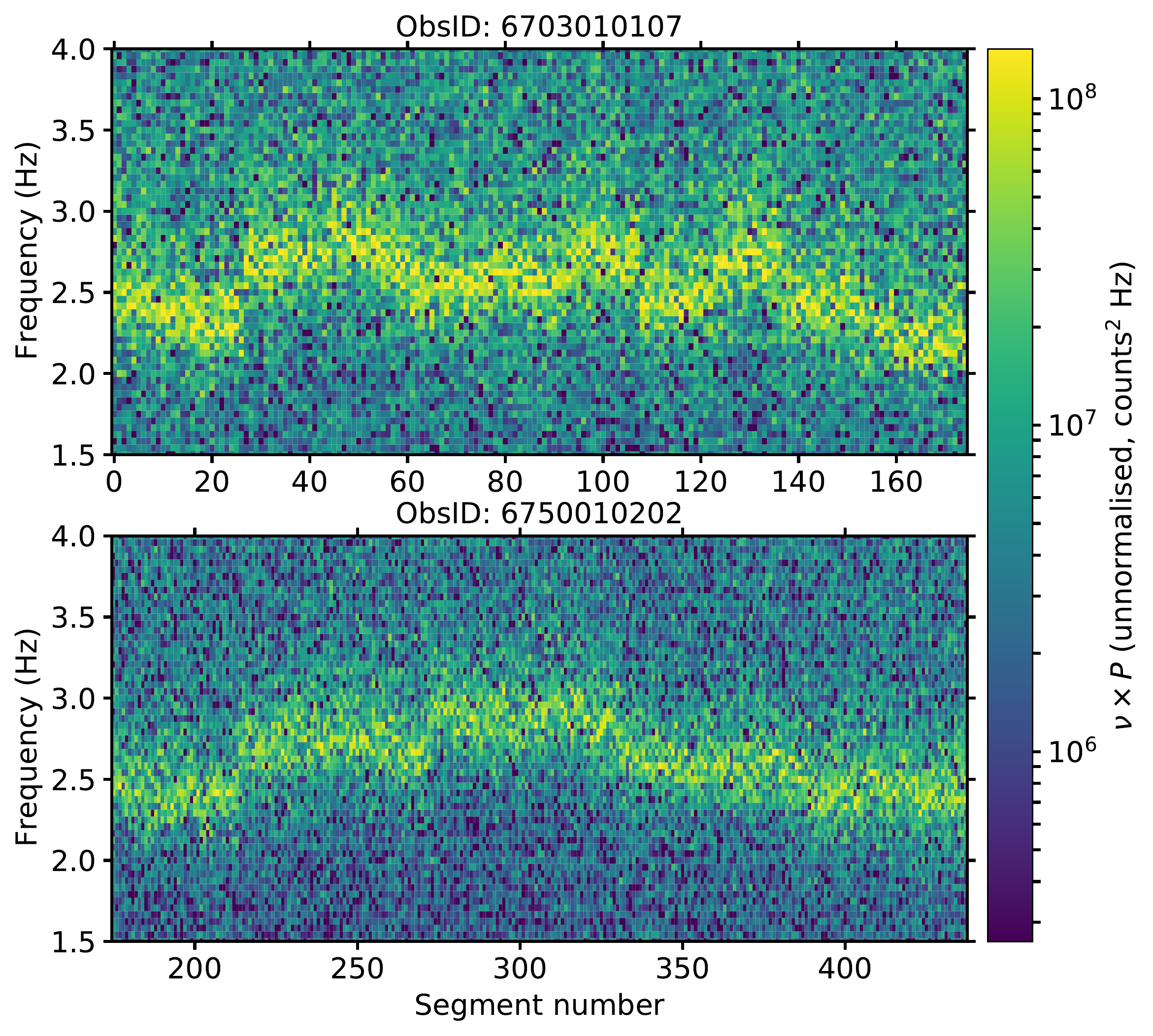}
    \vspace{-2mm}
    \caption{Dynamical power spectrum of the NICER observation of Swift~J1727.8$-$1613, with a QPO at \(2.54\) Hz. This observation combines two ObsIDs: the upper panel shows ObsID 6703010107, and the lower panel shows ObsID 6750010202. 
    }
    \label{fig:dynPDS}
\end{figure}

Fig.~\ref{fig:dynPDS} shows the dynamical power spectrum of the NICER observation of Swift~J1727.8$-$1613 with a mean QPO frequency of \(2.54\) Hz. The QPO frequency varies between $\sim$2.0 and $\sim$3.0 Hz during the observation. As a result, the bicoherence map shows the elongated feature around the QPO frequency pair in Fig.~\ref{fig:bic_bis_bip_J1727_2.67Hz}.


\bsp	
\label{lastpage}
\end{document}